\RequirePackage{fix-cm}
\documentclass{article}
\usepackage{graphicx}
\usepackage{amssymb}
\usepackage{hyperref}

\usepackage{listings}
\usepackage{pgfplots}
\pgfplotsset{compat=1.17}

\usepackage{xcolor}

\newcommand{\DCDP}{\textbf{CDPC}}
\newcommand{\GDCOP}{\textbf{GCOPC}}
\newcommand{\SCP}{\textbf{SC}}

\definecolor{darkgreen}{RGB}{0, 150, 0}

\newcounter{cdobleimpl}
\def\thecdobleimpl{\ifnum\value{cdobleimpl}=1 $\Longrightarrow$:\ \else $\Longleftarrow$:\ \fi}

\newenvironment{proof}{\par\textbf{\textit{Proof.}}\ }{\qed}

\def\squareforqed{\hbox{\rlap{$\sqcap$}$\sqcup$}}
\def\qed{\ifmmode\squareforqed\else{\unskip\nobreak\hfil
\penalty50\hskip1em\null\nobreak\hfil\squareforqed
\parfillskip=0pt\finalhyphendemerits=0\endgraf}\fi}

\newtheorem{theorem}{\bfseries Theorem}
\newtheorem{definition}{\bfseries Definition}
\newtheorem{proposition}{\bfseries Proposition}
\newtheorem{corollary}{\bfseries Corollary}
\newtheorem{lemma}{\bfseries Lemma}
\newtheorem{example}{\bfseries Example}

\newcommand{\ismcomment}[1]{}
\newcommand{\bdfn}{\begin{definition} \begin{rm}}
\newcommand{\edfn}{\end{rm}$ $\qed \end{definition}}
\newcommand{\bthm}{\begin{theorem} \begin{rm}}
\newcommand{\ethm}{\end{rm}$ $\qed \end{theorem}}
\newcommand{\bprop}{\begin{proposition} \begin{rm}}
\newcommand{\eprop}{\end{rm}\qed\end{proposition}}
\newcommand{\bcor}{\begin{corollary}\begin{rm}}
\newcommand{\ecor}{\end{rm} \end{corollary}}
\newcommand{\blem}{\begin{lemma} \begin{rm}}
\newcommand{\elem}{\end{rm}\qed\end{lemma}}
\newcommand{\bfact}{\begin{fact} \begin{rm}}
\newcommand{\efact}{\end{rm} \end{fact}}
\newcommand{\bex}{\begin{example} \begin{rm}}
\newcommand{\eex}{\end{rm}$ $\qed  \end{example}}

\newcommand{\bprf}{\begin{proof}}
\newcommand{\eprf}{\end{proof}}

\begin{document}

%
%
%
%
%

\title{Complexity analysis and practical resolution of the data classification problem with private characteristics\thanks{This paper was published in Complex \& Intelligent Systems. The present version is the author's accepted manuscript. This work has been partially supported by Spanish projects PID2019-108528RB-C22 and PID2023-149943OB-I00.
}}

\author{David Pantoja$^{1}$,
        Ismael Rodríguez$^{1,2}$,\\
        Fernando Rubio$^{1,2}$,
        Clara Segura$^{1}$,
\thanks{$^1$
Dpto. Sistemas Inform\'aticos y Computaci\'on.
Facultad de Inform\'atica.
Universidad Complutense de Madrid. 28040 Madrid, Spain.}
\thanks{$^2$Instituto de Tecnolog{\'\i}as del Conocimiento.}
\thanks{E-mail: {\tt dpantoja@ucm.es}, {\tt isrodrig@sip.ucm.es}, {\tt fernando@sip.ucm.es}, {\tt clsegura@ucm.es}%
}}

\date{}

\maketitle

\begin{abstract}

In this work we analyze the problem of, given the probability distribution of a population, questioning an unknown individual that is representative of the distribution so that our uncertainty about certain characteristics is significantly reduced ---but the uncertainty about others, deemed private or sensitive, is not.
Thus, the goal of the problem is extracting
information being relevant to a legitimate purpose while preserving the privacy of individuals,
which is crucial to enable non-intrusive selection processes in several areas.
For instance, it is essential in the design of non-discriminatory personnel selection, promotion, and layoff processes in companies and institutions; in the retrieval of customer information being relevant to the service provided by a company (and no more); in certifications not revealing sensitive industrial information being irrelevant for the certification itself; etc.
Interactive questioning processes are constructed for this purpose, which requires generalizing the notion of {\it decision trees} to account the amount of desired and undesired information retrieved for each branch of the plan.
Our findings about this problem are both theoretical and practical: on the one hand, we prove its NP-completeness by a reduction from the Set Cover problem; and on the other hand, given this intractability, we provide heuristic solutions to find reasonable solutions in affordable time. In particular, a greedy algorithm and two genetic algorithms are presented.
Our experiments indicate that the best results are obtained using a genetic algorithm reinforced with a greedy strategy.
\end{abstract}


\section{Introduction}
\label{intro}

Personal information is released both voluntarily and inadvertently during our interactions with others.
Since there is a common perception of the most common characteristics of individuals of a specific age, gender, ethnicity, culture, education, wealth, etc., the answers of individuals about apparently innocent questions regarding their situation or habits may reveal, all together, that they belong to a specific group with high probability, which might be a source of discrimination~\cite{purkiss2006implicit,macan2011actions,kroll2016discrimination}. Even though releasing some information is necessary to enable a service or process (e.g. a medical test or a job interview), certain released information may lead to the discovery of private characteristics which are irrelevant to the process under consideration ---or at least, provide little relevant information about it, to the extent that enough relevant information can also be achieved in an alternative way providing negligible information about sensitive issues.
Hence, it is crucial to design questioning processes such that, in any case, the answers will provide sufficient information regarding the relevant issue under consideration (e.g., the aptitude of a person to a job) without revealing excessive information about other characteristics considered private (e.g., the religion or the sexual orientation).

Technically, we wish to significantly reduce the uncertainty\footnote{Note that there are different types of uncertainty. These can be internal or external, where internal uncertainties are simply due to a lack of knowledge (either complete or partial) and external uncertainties are irreducible, such as those due to random effects. Uncertainty can also be classified as parametric or non-parametric. Non-parametric uncertainty is due to lack of correctness of the model, whereas parametric uncertainty refers to lack of precision in the estimation of the parameters of a model assumed to be correct. In our case, we will deal with internal and parametric uncertainty: no randomness in the applicant's answers is assumed, and the applicant’s features are assumed to follow the given population statistical dataset, although we do not know them yet.} regarding the former issues (e.g., making the answers reveal whether the candidate is competent to the job with more than 85\% or less than 15\% probability in all cases) without significantly reducing the uncertainty about the latter ones (e.g., if 30\% of the general population has certain sensitive private characteristic, then we wish that, for any possible answers to our questions plan, the probability of having this characteristic conditioned to the provided answers is neither lower than 20\% nor bigger than 40\%).
Note that we cannot achieve this by changing the common knowledge basis, because this information is either well known for anyone with enough education or has been released in public polls (e.g. it is well known that followers of some religions do not eat pork). Thus, we consider the problem of designing {\it adaptive} questionnaires (i.e. the next question to be posed may depend on previous answers) inferring as {\it much} as possible about the target characteristics while inferring as {\it least} as possible regarding those characteristics deemed private.

One of the applications of our problem is the design of non-discriminatory personnel selection, promotion, or layoff processes in companies or institutions, a problem of great interest nowadays~\cite{dattner2019legal,kodiyan2019overview}.
Consider a scenario in which a human resources manager is tasked with designing a selection process that consists of a series of questions aimed at extracting the maximum amount of job-relevant information. This information is intended to predict the future performance of each respondent, thereby determining their suitability for the position.
Simultaneously, the manager wants to ensure that no respondent's information considered private, such as, e.g. nationality, religion, or sexual orientation, can be inferred from the answers to the questions. We assume that the manager has access to a sizeable dataset that contains job performance information on individuals, and the individual's characteristics annotated in the dataset are precisely the questions the manager may or may not ask about.
This dataset may come from historical data of company employees or from common knowledge on general personnel performance.
The likelihood that the respondent has certain characteristics will be assumed to be the same as the ratio of individuals with these characteristics in the dataset. Hence, sometimes individuals in the dataset will be referred to as {\it candidates}.

For example, suppose that the percentage of homosexuals in the dataset population is 10\%. Then, for any given respondent, their answers to the questions in the designed interview must allow us to determine their fitness in such a way that the percentage of homosexuals in the \textit{subset of people who would answer in the same way} is not substantially higher or lower than 10\%. In this way, we ensure that this additional information cannot be used with discriminatory purposes, as the probability of having or not having a potentially discriminatory characteristic is more or less the same as in the dataset (which is assumed to be, in turn, representative of potential candidates or of the general population). That is, the uncertainty about the potentially discriminatory characteristic remains more or less the same as before answering any question.

It is possible to infer private information from questions that at first glance may seem irrelevant to it. Many questions can infer private information when asked in conjunction with certain other questions, even though when asked individually they may not infer any private information. Alternatively, there are questions that individually may infer private information, but when asked after certain information has been obtained, may not infer such private information.
For example, let us consider a society where vegetarianism does not correlate with following some religion, and let us suppose that we already know that the respondent does not eat chicken. Then, asking afterward whether they eat pork will most likely not reveal their religion, since probably the respondent is vegetarian or does not like chicken.
Because of this, it is of great interest to design the interview so that responses to previously asked questions influence the selection of subsequent questions. In this sense, the interviews are going to be \textit{interactive}.

The problem under consideration can also be used in other different contexts.
For instance, it is essentially the same in the design of user classification methods for a service, such as a web page, when we want to ensure that they do not extract sensitive information about the user which is irrelevant to the service being provided.
This can be very valuable in the automated creation of user profiles; e.g. profiles used to personalize advertising.
These profiles are created from information provided by users through, among others, cookies.
The creation of these profiles must comply with privacy and personal data protection laws, so designing methods to classify web page users that do not extract sensitive information directly (such as through forms) or indirectly (such as through browsing history and click rate) is of great importance nowadays~\cite{ullah2020privacy,estrada2017online}.
This problem can also be applied in the design of certification processes of products or services in such a way that private data, such as industrial secrets, are not revealed.

We will show that the problem is NP-complete. Although our interviews can be seen as {\it decision trees}, the NP-hardness of their construction in our problem is not a consequence of the NP-hardness of constructing minimal decision trees~\cite{optimalBinaryDecisionTrees}, as our problem will be NP-hard even  though the size or deepness of the trees is not limited ---a requirement which is necessary for the NP-hardness of constructing decision trees which {\it just} reduce the uncertainty about some characteristic.\footnote{Note that if no size or length limit is considered in that case, then we only have to analyze the tree posing all questions in all branches, which makes the problem trivial.} On the contrary, in our case the difficulty will come from the fact that additional questions reduce the uncertainty of both our target characteristics {\it and} the sensitive characteristics, so both goals are  contradictory in general.

Given the NP-hardness of the problem, it is computationally infeasible to provide exact solutions efficiently unless $P = NP$. However, we will also show that it is possible to find reasonable solutions in affordable time by using heuristic algorithms. In particular, we will introduce a greedy algorithm to compute suboptimal solutions, and we will also introduce genetic algorithms to solve the same problem.

To the best of our knowledge, the problem posed in this paper is the first one studying the construction of decision trees where the goal is reducing the uncertainty about some features while keeping the uncertainty about other features as intact as possible.
As suggested above, our problem differs from {\it decision tree construction} problems in that the difficulty arises from two contradictory goals rather than from any size limit.
Besides, as we will see next in Section~\ref{sec:related work}, our problem will differ from {\it machine learning data classification} techniques in that we must {\it avoid} accurately classifying  some label of the individual while we try to classify another one;
it also will differ from {\it private learning} techniques in that the original information cannot be pruned or obfuscated, because it is assumed to be well-known;
and it will not concern {\it randomize response and negative survey} methods, because our goal will be obtaining information about one individual rather than about the general population.
The main contributions of the paper are the following:
(1) formally defining the data classification problem with private characteristics;
(2) formally proving the NP-completeness of the problem;
(3) providing algorithms to solve the problem in practical situations;
and (4) conducting an experimental study accompanied by a statistical analysis of the results.

The rest of the paper is structured as follows. Next we discuss related works, explaining the differences between our problem an others that are somehow similar to ours. Then, Section~\ref{sec:definition} is devoted to formally introduce and define the problem under consideration. Afterwards, in Section~\ref{sec:complexity analysis} we sketch the NP-completeness proof of our problem. Section~\ref{sec:practical resolution} introduces different algorithms to deal with our problem, and
these algorithms are evaluated using concrete benchmarks.
Finally, Section~\ref{sec:conclusions} presents our conclusions and lines for future work. The interested reader can also find in Appendix~\ref{sec:appendix proof} the detailed formal proof of the NP-completeness of our problem.

Let us point out that Section~\ref{sec:complexity analysis} (along with the corresponding formal proof presented in Appendix~\ref{sec:appendix proof}) is independent of the algorithms and experiments discussed subsequently. Proving the time complexity of the problem is relevant to justify the use of heuristics to solve it (see e.g.~\cite{galiana2023stop,godoy2024voting}), but a practitioner reader can safely skip the section and appendix dealing with the computational complexity, and focus on the practical resolution of the problem.

\subsection{Related work}
\label{sec:related work}

To the best of our knowledge, the problem introduced and analyzed in this paper has not been previously investigated in the literature.
Next we mention several problems which are somehow related and explain how they differ from our study. They can be classified into three groups.

\paragraph{Decision trees.} Given a finite set of testable objects $X$, a finite set of tests $T$, and $w\in \mathbb{N}$, where for every test $t\in T$ and object $x\in X$ the value of $t(x)$ is either true or false, the decision tree problem consists in determining if there exists an identification procedure (in the form of a binary decision tree) able to completely identify any given object in the set and whose external path length (sum of the lengths of the paths from the leaves of the tree to the root) is less than $w$. The decision tree problem is known to be NP-complete~\cite{optimalBinaryDecisionTrees}.
Similarly, several variants of the problem of finding out whether it is possible to distinguish the correctness of an \textit{implementation under test} (IUT) by means of adaptive tests have been studied when the set of possible correct and incorrect behaviors is defined extensionally, i.e.\ case by case~\cite{datdp}. Instead of limiting the external path length, these variants require the depth of the tree to be limited. Additionally, the tests can have non-binary answers, and some variants allow non-deterministic behaviors from the IUTs.

Our problem differs from the problems studied in these works in that we are not interested in limiting the number of questions (or tests) that can be asked, but in limiting the amount of private information these trees allow to infer in any case. By posing more questions, more desired {\it and} undesired information is gained in general, and care must be taken to make these questions highly reduce the uncertainty about the former while not doing so about the latter in all branches of the questioning tree.
Hence, the tree size (measured in depth or in number of nodes) is not restricted at all by the problem definition, but by the necessity to avoid branches revealing too much undesired information. Indeed, it is the subtle interaction between two (generally) contradictory forces what will make trees have balanced size in general.\footnote{Note that long branches are still possible ---for instance, if they pose questions which are more or less orthogonal to the characteristics not to be learned.} The absence of any explicit size limitation means that our problem is not a generalization of any of the previous problems, so a specific NP-hardness proof is needed ---and since it is not a particularization either, a specific proof of ownership to NP is needed as well.

\paragraph{Machine Learning data classification techniques.} Works in this field apply machine learning for data classification~\cite{1SupervisedLearningInDataClassification}. The classifier is trained with a dataset that contains information on individuals in the form of tuples of data. These tuples are labeled, so supervised machine learning is considered in this case.
Some classifiers use the information collected to build decision trees. To avoid overfitting (learning the training data too well and not performing well on other datasets), it is common to prune subtrees.
For instance, different pruning techniques are presented in~\cite{1PruningDataClassification}, also analyzing their complexity.
Note that, in our work, we do not consider machine learning approaches to data classification, but we analyze the time complexity of a non-random, non-approximation~\cite{aproximabilidad,munoz2021evaluating} of this problem and next develop heuristic methods for it.
More importantly, we are not only trying to predict the class label (which in our case will be the fitness to some goal), but we deal with attributes deemed sensitive (private characteristics) and we impose restrictions on the amount of information we can infer about them while trying to predict the class label.

\paragraph{Private learning.} The works in this field discuss privacy-preserving data classification methods \cite{3DecisionTreeClassificationWithDifferentialPrivacy,3AnonymizingClassificationDataPrivacyPreservation,3PrivacyPreservingClassificationOverDifferentiallyPrivateData}. These methods anonymize the information of the individuals in the dataset that is being utilized to train the classifier. Many datasets contain information in the form of tuples of data, like $\mathit{(David, Spain,}$ $\mathit{9, August, 2001)}$. However, a naive approach that simply removes the identifiers from the tuple (in our example, only the name) may be insufficient to guarantee the anonymity of the individuals in the dataset. One popular anonymization technique is \textit{k-anonymization}, which ensures that each individual in a dataset is not distinguishable from at least $k$ other individuals~\cite{caballero2017anticipating}.
Unlike these methods, in our problem we are not trying to anonymize the information of the individuals in the dataset, because (a) the dataset will be assumed to refer to groups of people (or stereotypical types of individuals) rather than specific individuals, and (b) it will be assumed to denote common demographics knowledge which is usually known by specialists in some field or can be extracted from public polls ---for instance, it may show general differences in consumption habits by age, gender, and wealth. Since the dataset is commonly known, it {\it cannot} be modified, pruned, or obfuscated by any means.
Instead, we are trying to obtain relevant information about the individual being questioned (assuming the likelihood that the individual has any combination of characteristics is reflected in the dataset) without inferring characteristics deemed private about such individual.

\paragraph{Randomized response techniques and negative survey methods.} Both randomized response techniques and negative survey methods are approaches to constructing surveys that ensure that no private information is inferred from the respondent. Randomized response techniques require questions with binary answers, while negative survey methods require questions with more than two answers.

Surveys that follow the randomized response techniques~\cite{warner1965randomized} ask the respondents for their membership in one of two mutually exclusive and exhaustive groups (that is, if they are currently residing in France or if they are \textit{not} currently residing in France). The question is selected from the two possible options using a randomizer, and the selection is only disclosed to the respondent. Thus, the respondents do not reveal to what group they belong to, but the characteristics of the randomizer and the proportions of the \textit{Yes} and \textit{No} answers given are sufficient to estimate the proportion of population in each group. Alternatively, the less restrictive \textit{paired-alternative}~\cite{greenberg1969unrelated} variant of this technique pairs sensitive questions with unrelated questions.

On the other hand, negative survey methods~\cite{esponda2006negative} utilize surveys in which the respondent is asked for a category they do \textit{not} belong to. For example, instead of asking for the political party they are going to vote in the upcoming elections, the respondents are asked for a political party they are not going to vote. Similarly to randomized response techniques, these methods are used to obtain information about the general population, without revealing information about the respondents.

It is important to emphasize that the purpose of randomized response techniques and negative survey methods is to obtain information regarding the population as a whole, rather than about individual respondents. Therefore, they are appropriate for conducting surveys instead of interviews and are fundamentally different from our approach. In our problem it is essential to gather specific information about each individual respondent.

\section{Definition of the problem}
\label{sec:definition}

In this section, we introduce the problem under consideration. First, in order to illustrate the main aspects of our problem, a simple and informal example is presented.
After that, the formal definition of the problem is given.

\subsection{Informal example of an instance of the problem}
\label{sec:informal example}

Returning to the example of a human resources manager of a company, let us assume that the manager has a dataset with information about five characteristics of hundreds of employees of the company. These characteristics are nationality (\textit{local/foreign}), programming language skills (\textit{high/low}), work experience (\textit{yes/no}), education level (\textit{master's/bachelor's/none}) and fitness (\textit{fit/unfit}). Fitness can be, for example, a measure of the employee's performance in the company.

The manager has to determine whether a candidate would perform well or not {\it without} inferring their nationality, as it will be deemed private information. The manager is going to use the information in the dataset to design an interactive interview that does not infer private information, but predicts with as much certainty as possible the future performance of the respondent. The available data, which were composed only to serve as an example, can be seen in Table~\ref{tab:informal example problem}.

\begin{table}
    \caption{Example of the data classification problem with private characteristics}
    \label{tab:informal example problem}
    {\scriptsize
    \begin{tabular}{|c|c|c|c|c|c|c|} \hline
            &\textbf{Nationality}&  \textbf{Programming}&  \textbf{Experience}&  \textbf{Education}&\textbf{Fitness}&\textbf{Quantity}\\ \hline
            1&local&  High&  Yes&  Master's&Fit&100\\ \hline
            2&local&  Low&  Yes&  Master's&Unfit&100\\ \hline
            3&local&  Low&  No&  Master's&Fit&100\\ \hline
            4&local&  Low&  No&  Bachelor's&Unfit&100\\ \hline
  5&local& Low& No& None&Unfit&50\\\hline
            6&foreign&  High&  Yes&  Bachelor's&Fit&100\\ \hline
            7&foreign&  Low&  Yes&  Bachelor's&Unfit&100\\ \hline
            8&foreign&  High&  No&  Master's&Fit&100\\ \hline
  9&foreign& High& No& Bachelor's&Unfit&100\\ \hline
  10&foreign& High& No& None&Unfit&50\\\hline
    \end{tabular}
    }
\end{table}

The column \textit{Quantity} indicates the number of employees who respond in the way described by each row. Note that not all possible combinations of characteristics are explicitly defined; combinations that do not correspond to any employee are not stored in the dataset.

The interview, i.e. the problem solution we are looking for,
is a tree. The internal nodes represent the questions and
the arcs going down from them represent each possible answer to its associated question. The leaf nodes represent the last states of the interview.

\begin{figure}
    \includegraphics[width=1\linewidth]{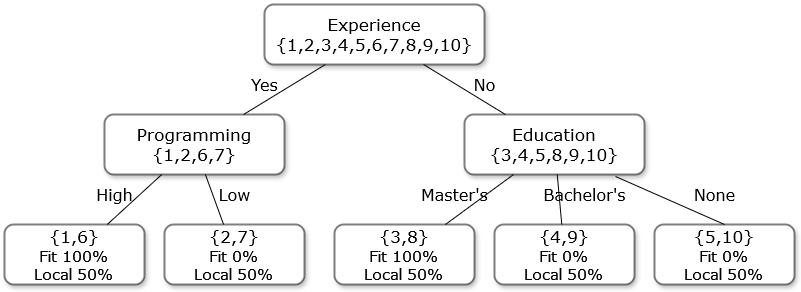}
    \caption{Example of an interactive interview}
    \label{fig:ejemplo entrevista interactiva}
\end{figure}

The tree in Figure~\ref{fig:ejemplo entrevista interactiva} is an example of an interactive interview that the human resources manager can design. The interviewer will start by asking about the respondent's experience. If the respondent responds that they have experience, then the interviewer will ask about their level of programming. If they have no experience, then the manager will ask about their level of education. We can see that, regardless of the respondent's answer to the first and second question, all of the company's employees who would have answered in that way are equally fit, and exactly half of them are local (which is the same as in the whole dataset). Assuming that the population of possible respondents is similar to that of the employees, we can deduce (in this case, with no margin of error) the fitness of any given respondent without having inferred the information considered private in this case.

Note that not all interviews would yield equally effective results. Asking about the respondent's level of programming before asking about their work experience would infer private information about their nationality, since locals have a worse level of programming than foreigners in our example. Similarly, asking about their education level instead of their programming level after the respondent says that they have work experience would fail to distinguish the fit individuals from the unfit ones. This showcases the importance of choosing the right interview, and hints at the computational difficulty of this process.

We are going to study the construction of privacy-preserving interviews as a decision problem. Thus, there will exist instances of the problem for which it is impossible to design interviews that satisfy their given constraints. If there is an extreme correlation between the target characteristic and the private characteristic, then it will be impossible to gain substantial knowledge about the former without acquiring an extensive understanding of the latter. Hence, for some demanding thresholds of minimum knowledge to be reached about the former and maximum allowed knowledge about the latter, sometimes the answer will be \textit{No}.

This does not mean that highly correlated problem instances can never be solved, particularly if the required thresholds are not very demanding, because in these cases the thresholds can be respected by cleverly exploiting the small differences between both characteristics. For instance, if the target characteristic and the private characteristic coincide in $7/8$ of the cases (or are exactly the opposite in $7/8$ of the cases, which is an equally difficult situation), then any room to obtain more information about the former than about the latter lies precisely in exploiting the $1/8$ of cases where the answers are different. Let us suppose each bit in the vector $01110001$ denotes whether the target characteristic \textit{t} is present or absent for individuals who respond to the questions \textit{x}, \textit{y}, \textit{z} in a specific manner: \textit{No}, \textit{No}, \textit{No} for the first bit; \textit{No}, \textit{No}, \textit{Yes} for the second; etc; and \textit{Yes}, \textit{Yes}, \textit{Yes} for the eighth. Similarly, let us suppose that $11110001$ is the respective bit vector for the private characteristic \textit{f}. We can see that the characteristics \textit{t} and \textit{f} coincide in $7/8$ of the cases, and as such this is a highly correlated scenario. We will assume that the proportion of individuals that respond in each of these eight ways to the questions is evenly distributed (that is, $1/8$ of the candidates respond in each of the 8 possible ways to the questions \textit{x}, \textit{y}, \textit{z}).

In order not to infer more information than permitted by the instance's constraints about \textit{f}, asking the question \textit{x} is incorrect, because if the respondent answers \textit{No} then we will know that he has the private characteristic \textit{f}. In this case, it is cleverer to ask the question \textit{y}: responding \textit{Yes} to \textit{y} means 75\% probability of having \textit{f}, whereas responding \textit{No} means 50\% probability. Note that these probabilities do not highly deviate from the \textit{a priori} probability of having \textit{f}, which is 62,5\% ($5/8$). Still, the question \textit{y} provides valuable additional information about characteristic \textit{t}: responding \textit{Yes} means 75\% of having \textit{t}, whereas responding \textit{No} means 25\%. Actually, these probabilities deviate from the \textit{a priori} probability of having \textit{t} (50\%) more than the deviation produced from the \textit{a priori} knowledge of \textit{f} by making the same question \textit{y}. Hence, by asking question \textit{y} we gain more information about the target characteristic \textit{t} than about the private characteristic \textit{f} —--despite the highly correlated instance of the problem.

\subsection{Formal definition of the data classification problem with private characteristics}
\label{sec:formal definition}

In the definition of the problem, we will call $\mathit{Answer}$ the finite set of all the answers that the respondents can give to the questions. For example, if the candidates can only answer \textit{Yes} or \textit{No} to the questions, then $\mathit{Answer=\{Yes, No\}}$. If questions have different answers, then the set $\mathit{Answer}$ will be the union of all possible answers to the questions.

Similarly, $\mathit{Question}$ is the finite set of all questions that can be asked in the interview. It should be noted that the questions that can be asked and the characteristics of the candidates are analogous concepts; each question asks explicitly for a single characteristic and all characteristics have an associated question that asks explicitly for that characteristic. For example, if one characteristic is age, then there is exactly one question that asks directly about age. Hereafter we will use the name of the characteristic (\textit{Age}) instead of the interrogative sentence (\textit{How old are you?}).

Finally, we need to specify how we are going to represent the candidates. It will not be necessary to define individually candidates that are equivalent to others. A \textit{candidate type} will be a group of candidates who answer all the questions in the same way. Candidate types will be total functions $\mathit{Question}\rightarrow \mathit{Answer}$. Each function can be considered as a way of answering questions.


For example, if the candidates of type $c$ are 30 years old and female, then $c(age)=30$ and $f(gender)=female$. Instead of defining the fitness of each candidate separately, we will define the fitness of each candidate type and the number of candidates of each type. We will say that a candidate type \textit{represents} a candidate if that candidate answers the questions as specified by the candidate type.

We will assume that the respondent is representative of the probability distribution of the candidates in the dataset, that is, the likelihood of facing a respondent represented by a specific candidate type is the same as the ratio of the individuals of that type in the dataset.
Note that our goal is not to ask questions until the candidate type is fully determined, but rather to ask questions in such a way that, regardless of the respondent's candidate type, the uncertainty about a certain characteristic is highly reduced while it is not for the private characteristics.
Hence, questions will (generally) stop before the candidate type of the respondent has been fully determined (i.e. several candidate types are still possible).
When we say that e.g. the respondent has a 50\% chance of being fit or of having any other characteristic, we mean that 50\% of the candidates that would have answered the questions asked until that point in the same way as the respondent have the characteristic.
The necessary assumption that the respondent is as in the dataset implies that the dataset is complete enough so that the respondent always matches a candidate type in it
(i.e. there exists a candidate type that represents the respondent).\footnote{Due to the lack of enough information concerning very unlikely cases, in practical situations a theoretically-impossible respondent not matching any candidate type could be faced. This means handling a respondent assumed to have $0$ probability according to the underlying knowledge (i.e. the dataset), so no information can be provided according to that knowledge. Consequently, the questionnaire tree to be constructed is not expected to be any useful in this aberrant case. Given its negligible probability, its effect on the overall tree usefulness is negligible as well. In technical terms, the consequent uncertainty of the lack of sufficient information in the dataset is referred to as \textit{interpolation uncertainty}~\cite{uncertainty}.}

\begin{definition} {\it (\DCDP)} An instance of the {\it data classification decision problem with private characteristics} (\DCDP) is a tuple $(f, g, P, a, b, x, y)$, where, assuming $n\in \mathbb{N}$ is the number of questions (and characteristics):

\begin{itemize}
    \item $f:(\mathit{Question}\rightarrow \mathit{Answer}) \rightharpoonup \{\top, \bot\}$ is a partial fitness function that evaluates candidate types. Since the problem is non-exhaustive, the functions $c:\mathit{Question}\rightarrow \mathit{Answer}$ that do not represent any candidates are not in the domain of $f$. The image of $f$ is ${\top, \bot}$, where $\top$ indicates that the candidate type is fit and $\bot$ that it is not fit.

    \item $g:(\mathit{Question}\rightarrow \mathit{Answer})\rightharpoonup \mathbb{N}$ is a partial function that returns, given a candidate type, the number of candidates that can respond in that way. Candidate types that do not represent any candidate are not in the domain of $g$.

    \item $P\subseteq \mathit{Question}$ is the set of private characteristics.

    \item $a, b\in \mathbb{Q}$, where $0\leq a\leq b\leq 1$, restrict the amount of private information that can be inferred. We must ensure that, for each private characteristic, the ratio of the population that would respond a given specific answer to that question
    will remain between $a$ and $b$ after each response (this particular answer will be referred to as $1$ for the sake of notation simplicity, although it may be any particular answer, e.g. ``Yes''). For example, if $a=0.2$ and $b=0.7$ then, during the interview, the percentage of the candidates who have not been ruled out who would answer the private question “Are you under 30 years old?” with “Yes” will have to be maintained between $20$ and $70$ in all possible cases. Otherwise, we will consider that private information about the respondent's age has been inferred. We will refer to candidates that answer 1 to the question associated to a private characteristic as candidates \textit{with} that private characteristic.\footnote{Note that, if three or more answers are possible, then defining limits only for one of them (answer $1$) indirectly limits the remaining ones
    ---in particular, only the {\it sum} of their probabilities is restricted by the limits of answer $1$. Despite being simple and unexpressive, this limited format will be enough to make the problem NP-hard (in fact, just binary answers will be used to prove that). A generalization of this problem, given in Definition~\ref{def:definition generic problem}, will enable a much more natural and customized definition of private limits ---as well as candidate types which are not {\it fully} fit or unfit.}
    \item $x,y\in \mathbb{Q}$, with $0\leq x\leq y\leq 1$, indicate how sure we should be about our prediction of the fitness of the respondent.
    All possible paths of the tree will have to narrow down the set of matching candidates (i.e. the set of candidates that would answer in the same way as the respondent to the questions in the path of the tree) to a subset whose ratio of fit candidates is less than $x$ or greater than $y$. For example, if $x=0.2$ and $y=0.8$, then in all cases we will need to narrow down the population to a subset that is sufficiently unfit (with 0\% to 20\% of fit candidates) or fit (with 80\% to 100\% of fit candidates).
\end{itemize}
\end{definition}

Obviously, the problem instance does not include which candidate is the respondent, as the goal of the tree is being able to handle {\it all} possible candidates as specified.

With each question we ask the respondent, we are essentially dividing the \textit{set of candidates who match the respondent} into two; those candidates who would give the same answer as the respondent and those who give any other possible answer (except in the case where all remaining candidates would answer that question the same way). Although from the interviewer's point of view the population is divided in two for each answer of the respondent, the tree of questions and answers is not binary. Each question can have more than two answers, and the interactive interview that solveslves the problem must be able to deal with all options.

We will not allow repeated questions in the same path of an interview because, assuming consistency in the answers, repeated questions do not provide any further separation and are redundant. Therefore, in the worst case the depth of the tree will be the number of questions in the instance of the problem.

An instance of \DCDP\ is \textit{positive}, i.e. the answer of \DCDP\ for that instance is $\top$, if there exists an interactive interview (tree) that ensures, for any possible respondent, that we can predict their fitness with enough certainty (according to $x$ and $y$) and that we do not infer more private information than permitted (according to $a$ and $b$), and \textit{negative}, i.e. the answer of \DCDP\ for that instance is $\bot$, in any other case.

It could be argued the \DCDP\ definition has too little flexibility to denote practical situations. For example, our definition requires the fitness of the candidate types to be either $\top$ or $\bot$ (fit or unfit), but it might be interesting to allow candidate types to be, say, 70\% fit ---denoting that
70\% of the individuals in that type are fit and 30\% of them are unfit. Furthermore, the bounds $a$ and $b$ that restrict the amount of private information that can be inferred are shared for all private characteristics. Many practical situations may require definitions with more flexibility. The reason to leave the definition like this is that proving the hardness of a problem for its most limited definition is a more interesting classification, as hardness (e.g. NP-hardness) trivially propagates via generalization.
Thus, when the NP-hardness of a problem is being proved, it is preferable to analyze the complexity of the most restrictive variant of the problem, since more general variants immediately inherit the complexity of the more particular ones. Because of this, analyzing the complexity of \DCDP\ is of greater interest than studying the complexity of other more general and flexible variants of the problem. Despite this, we also define a variant of the problem that is more applicable in real situations. Actually, this more realistic variant will be the one used later in our experiments.


\begin{definition} {\it (\GDCOP)}
\label{def:definition generic problem}
An instance of the {\it generic data classification optimization problem with private characteristics} (\GDCOP) is a tuple $(f_g, g, P_g,$ $ k)$ that differs from instances of \DCDP\ in the image of the fitness function, the addition of a question limit, and the way of defining private characteristics:

\begin{itemize}
    \item $f_g:(\mathit{Question}\rightarrow \mathit{Answer}) \rightharpoonup \{q\ |\ q \in \mathbb{Q}, 0\leq q\leq 1\}$ is a partial fitness function that evaluates candidate types. As we mentioned before, $f_g$ allows the candidate types to have partial fitness (e.g. 70\% of the individuals represented by a type are fit).

    \item $k\in\mathbb{N}$ is the question limit $k$. It simply restricts the maximum number of questions in the interview (i.e. the depth of the tree) to be less than or equal to $k$.

    \item $P_g\subseteq \mathit{Question}\times \mathit{Answer}\times \{a\ |\ a\in \mathbb{Q}, 0\leq a\leq 1\} \times \{b\ |\ b\in \mathbb{Q},\ 0\leq b\leq 1\}$ is the set of tuples that defines the private characteristics. As aforementioned, having to use the same bounds $a$ and $b$ for all private characteristics might not be realistic in practical situations. \DCDP\ instance parameters $P$, $a$, and $b$ are replaced in \GDCOP\ by the set of tuples $P_g$, where each tuple $(r, p, a, b)$ represents a private answer $r$ (an answer that we consider sensitive) to a question $p$ (that inquires about a private characteristic) and the interval $[a,b]$ in which the ratio of candidates matching the respondent (i.e. the candidates that would answer the questions in the same way as the respondent) that would answer $r$ to the question $p$ should remain throughout the interview. For example, if $P_g=\{(age, <30, 0.1, 0.8), (age, >60, 0.4, 0.6), (Spanish, C1, 0.3, 0.7)\}$, then the ratios of candidates less than 30 years old, more than 60 years old, and whose level of Spanish is C1 should remain in the intervals $[0.1, 0.8]$, $[0.4, 0.6]$, and $[0.3, 0.7]$, respectively.

\end{itemize}

Additionally, \GDCOP\ differs from \DCDP\ in that, instead of determining sufficient high and low fitness levels to be passed in all cases (which were given in \DCDP\ by instance parameters $x$ and $y$) and ignoring the actual fitness values once they exceeded these  thresholds, this variant is an optimization problem. Therefore it aims to maximize the uncertainty reduction regarding the fitness (i.e. to get fitness values which are as close to $0$ or $1$ in all cases as possible) while keeping the privacy as desired. 
\end{definition}

Each path of the interview represents a series of questions and answers. The leaf nodes of the interview are associated to subsets of the initial population, resulting from having narrowed down the population to only the candidates that would respond in the way indicated by the path of the tree on which the node is located. Each leaf will have a certain fitness ratio, depending on the fitness of the candidates in its remaining population. The further away this ratio is from 0.5, the better we have distinguished the fit candidates from the unfit ones in the remaining population of the node. \GDCOP\ consists in finding out the interactive interview that distinguishes fit from unfit candidates the best in the worst case (i.e. on the interview leaf where the ratios of fit and unfit candidates are closer). Thus, in \GDCOP\ the interactive interview itself is returned, instead of $\top$ or $\bot$ as in the decision problem \DCDP.

\section{Complexity analysis}
\label{sec:complexity analysis}

In this section we informally introduce the main ideas of our NP-completeness proof. The reader interested in the detailed and formal proof is referred to Appendix~\ref{sec:appendix proof}, where the NP-completeness of the data classification problem with private characteristics (\DCDP) is formally proven.

The most relevant part of the proof is the definition of an appropriate polynomial reduction from another NP-complete problem (in our case, Set Cover). Thus, first we will explain how to define such reduction, and then we will illustrate it by using an example.

\subsection{Transformation from set cover to \DCDP}
\label{sec:transformation}



The {\it set cover decision problem} (\SCP) is a classical combinatorial optimization problem of great importance in the field of computational complexity. It is one of Karp's 21 NP-complete problems~\cite{Karp1972}. Given a tuple $(U, S, k)$, where the universe $U=\{u_1, ..., u_n\}$ is a set of elements, $S=\{S_1, ..., S_m\}$ is a set of subsets of $U$ whose union is the universe, $k\in \mathbb{N}$ and $k\leq |S|$, the set cover decision problem consists in determining if there exists a subset of $S$ of cardinality $k$ or less that covers the entire universe.

In Appendix~\ref{sec:appendix proof} it is proven that \DCDP\ is NP-hard by using a polynomial reduction from \SCP.
We need to define a polynomial algorithm that transforms instances of \SCP\ into instances of \DCDP. For the polynomial reduction to be correct, we need to ensure that, for any instance of \SCP, the output of \SCP\ for that instance is the same as the output of \DCDP\ for the instance resulting from the transformation of the original instance of \SCP.

As we have seen, an instance of \DCDP\ is a tuple $(U, S, k)$, where $U$ is a set of elements, $S\subseteq \mathcal{P}(U)$ and $k\in \mathbb{N}$, and an instance of \DCDP\ is a tuple $(f, g, P, a, b, x, y)$.

We will divide the transformation into two cases, depending on the value of $k$ in the instance of \SCP. If $k=|S|$ (including the case where $U=\emptyset$ and $k=0$), then the result of the instance is trivially $\top$ because we can select all sets in $S$ and $\bigcup_{S_i\in S} S_i = U$. Transforming these instances of \SCP\ into instances of \DCDP\ that are also positive is trivial. For example, we can use the instance of \DCDP\ defined in Section~\ref{sec:informal example}. For the rest of the study, we will assume that $k<|S|$. Furthermore, we will assume that $S$ does not contain the set $\emptyset$. If it does, then it could be trivially transformed into an equivalent \SCP\ instance by simply removing $\emptyset$ from $S$.

To transform instances of the set cover decision problem (where $k< |S|$), we first need to define the sets $\mathit{Question}$ and $\mathit{Answer}$. The set $\mathit{Question}$ contains the following elements:

\begin{itemize}
    \item Every set in $S$ (so $S\subset \mathit{Question}$). For example, if $\{1, 2, 3\}\in S$, then $\{1, 2, 3\}$ will be a possible question.
    \item An additional question which we will call \textit{private}. This question is necessary to represent the only private characteristic we will need in the \DCDP\ instance. Note that this question will not be part of the interview in any solution, since asking it would make impossible fulfill the restriction of not inferring private information. Nevertheless, this question is necessary because, according to our definition of the problem, every characteristic must be linked to a question.
\end{itemize}

On the other hand, it will be sufficient to allow binary responses. We define $\mathit{Answer}=\{1, 0\}$.

To define the fitness and quantity functions, that is $f:(\mathit{Question}\rightarrow \mathit{Answer}) \rightharpoonup \{\top, \bot\}$ and $g:(\mathit{Question}\rightarrow \mathit{Answer}) \rightharpoonup \mathbb{N}$, respectively, from an instance of \SCP, we will need to define the following types of candidates (represented with functions $\mathit{Question}\rightarrow \mathit{Answer}$), which will be the inputs of the functions $f$ and $g$:

\begin{itemize}
    \item A candidate type for every element $e\in U$. These candidates will answer $1$ to the questions corresponding to the sets to which they belong and $0$ to the others. For example, the candidate $c_3$ corresponding to element 3 will answer $1$ to question $\{1, 2, 3\}$ ($c_3(\{1, 2, 3\})=1$ because $3\in \{1, 2, 3\}$), $0$ to question $\{2\}$ ($c_3(\{2\})=0$ because $3 \notin \{2\}$), and $0$ to question \textit{private} ($c_3(private)=0$). We will refer to candidates belonging to these types as \textit{element}.

    \item A candidate type for each set $c\in S$. These candidates will answer $1$ to the question corresponding to their own set and $0$ to those corresponding to any other set. For example, the candidate $\{1, 2, 3\}$ will answer $1$ to question $\{1, 2, 3\}$ and $0$ to question $\{2, 4\}$. Additionally, they will answer $1$ to question \textit{private}. We will refer to candidates belonging to these types as \textit{set} and they are the only ones with the private characteristic.

    \item An additional candidate type, which we will call \textit{null}. These candidates answer $0$ to all questions.
\end{itemize}

The function $f$ will return $\top$ for \textit{null} candidates, as they will be the only fit ones. It will return $\bot$ for \textit{element} and \textit{set} candidates, and it will be undefined for other candidates.

There will be many more \textit{element} candidates than \textit{set} ones, and even more \textit{null} ones. The function $g$ will return $\Omega^2$ for \textit{null} candidates, $\Omega$ for \textit{element} ones, and $1$ for $set$ ones, where $\Omega\in \mathbb{N}$ is a value sufficiently high to guarantee the correctness of the transformation. Specifically, $\Omega=2*|U|*|S|$.

As we have mentioned, the only private characteristic is \textit{private}. Therefore, $P=\{private\}$.

In the initial population, the ratio of candidates with the private characteristic (those who would answer 1 to \textit{private}) is $|S| / (\Omega^2 + |U|*\Omega + |S|)$ because there are $|S|$ \textit{set} candidates that answer 1 to \textit{private}, $\Omega^2$ \textit{null} candidates and $|U|*\Omega$ \textit{element} candidates. The bounds $a, b\in \mathbb{Q}$ that define the minimum and maximum ratio of candidates that answer 1 to \textit{private} that can be reached without inferring more sensitive information than allowed will be $a=(|S| - k) / (|U|*\Omega + \Omega^2 + |S| - k)$ and $b=1$. Therefore, it will be acceptable for the entire population of candidates who match the respondent to respond 1. Finally, the bounds $x, y\in \mathbb{Q}$ that define the required fitness of the final population will be $x=0$ and $y=\Omega^2/(\Omega^2+|S|)$. Recall that in all cases we are required to reach a ratio of fit individuals sufficiently low (less than $x$) or high (more than $y$).

The number of input values for which the functions $f$ and $g$ are defined is polynomial with respect to the size of the input ($|U| + |S| + 1$) and we can calculate the value returned by these functions for any of these input values in constant time. $P$ can be constructed in constant time because it contains a single question. Finally, the cardinality of the sets $U$ and $S$ can be queried in linear time by simply traversing their elements, so the values of $a$, $b$, $x$ and $y$ can be computed in polynomial time. Since all the components of the tuple $(f, g, P, a, b, x, y)$ can be computed in polynomial time with respect to the size of $(U, S, k)$, the transformation we have defined is polynomial.

For the sake of clarity, next we present an example of the reduction. Let us remind that the formal proof is provided in Appendix~\ref{sec:appendix proof}.

\subsection{Example of the transformation to \DCDP}
\label{sec:transformation example}

In this section we illustrate how the transformation works by using an example. We will start from the following instance $(U, S, k)$ of the set cover decision problem:

\begin{lstlisting}

U = {1, 2, 3, 4}
S = { {1,2,4}, {1,3}, {2} }
k = 2

\end{lstlisting}

Recall that this decision problem consists of determining whether there is a subset of $S$ of cardinality less than or equal to $k$ such that the union of its elements is $U$. In our example, the result will be $\top$ because $|\{\{1,2,4\}, \{1,3\}\}|=2$ and $\{1,2,4\} \cup \{1,3\} = \{1,2,3,4\}$.

Table~\ref{tab:table_pcd_partial} illustrates the resulting instance of \DCDP. The rows represent the candidate types and the columns represent the answers to questions, the values given by the fitness function $f$ and the values given by the  quantity function $g$.

\begin{table}
    \begin{tabular}{|c|c|c|c|c|c|c|} \hline
           \textbf{Candidates}&  \multicolumn{4}{|c|}{\textbf{Questions}}&  \textbf{Fitness}& \textbf{Quantity}\\ \hline
  & \{1,2,4\}& \{1,3\}& \{2\}& \textit{private}& f&g\\ \hline
           1&  1&  1&  0&  0&  $\bot$& $\Omega$\\ \hline
           2&  1&  0&  1&  0&  $\bot$& $\Omega$\\ \hline
           3&  0&  1&  0&  0&  $\bot$& $\Omega$\\ \hline
           4&  1&  0&  0&  0&  $\bot$& $\Omega$\\ \hline
           \textit{null}&  0&  0&  0&  0&  $\top$& $\Omega^2$\\ \hline
           \{1,2,4\}&  1&  0&  0&  1&  $\bot$& 1\\ \hline
           \{1,3\}&  0&  1&  0&  1&  $\bot$& 1\\ \hline
           \{2\}&  0&  0&  1&  1&  $\bot$& 1\\ \hline
    \end{tabular}
    \caption{Resulting instance of \DCDP}
    \label{tab:table_pcd_partial}
\end{table}

For example, $f(c_1)=\bot$ and $g(c_1)=\Omega$, where $c_1$ is the candidate type $\mathit{Question}\rightarrow \mathit{Answer}$ corresponding to element 1 of the instance of \SCP, that is the one returning $c_1(\{1,2,4\})=1$, $c_1(\{1,3\})=1$, $c_1(\{2\})=0$, and $c_1(private)=0$.

All that remains is to define the values for $\Omega$, $a$, $b$, $x$ and $y$. $\Omega = 2*|U|*|S|=24$, $ a=(|S| - k) / (|U|*\Omega + \Omega^2 + |S| - k) = 0.0015$, $b=1$, $x=0$, and $y=\Omega^2/(\Omega^2+|S|) = 0.9948$.

The purpose of this transformation is to guarantee that there is a subset of $S$ that satisfies the instance of the set cover decision problem if and only if there exists an interactive interview that satisfies the instance of \DCDP.

Let a \textit{strategy} be a method of constructing the interactive interview where the selection of the next question is determined by the respondent's previous answers. Our strategy will be very simple: we will ask any of the questions corresponding to the solution of \SCP, i.e. the sets in the solution. If the respondent answers 1, then we will end the interview immediately, having concluded that they are unfit. If they answer 0, then we will move on to the next question and repeat this process, until the respondent answers 1 or there are no questions left corresponding to the solution of \SCP, in the latter case concluding that they are fit.

In our example, the interview following the aforementioned strategy consists of questions \{1,2,4\} and \{1,3\}. The order in which we ask the questions is not relevant. Next we check that, whoever the respondent is, we will be able to infer their fitness with enough accuracy (i.e. making its probability fall out of the $(x,y)$ range) without inferring more private information than allowed (i.e. keeping the probability of them answering $1$ to the private question within the $[a,b]$ range).

If the respondent is fit, then they can only be of type \textit{null}. Question \{1,2,4\} rules out candidate types 1, 2, 4, and \{1,2,4\}. A subsequent question \{1,3\} rules out types 1 (which was already ruled out), 3, and \{1,3\}. The only candidate types that have not been ruled out after both questions are \textit{null} and \{2\}. There are $\Omega^2$ candidates of the objective type (\textit{null}) and a single candidate of type \{2\}. Therefore, the fitness ratio of the remaining population is $\Omega^2/(\Omega^2+1) = 0.9983$, which is higher than the minimum required (as $y = 0.9948$).

On the other hand, if the respondent is unfit, i.e. not \textit{null}, then we need to consider two possible situations:

\begin{itemize}
    \item If the respondent is of type \{2\}, then they will have answered 0 to questions \{1,2,4\} and \{1,3\}. As the only candidate types that can answer 0 to both questions are \{2\} and \textit{null}, we are in a similar situation as above. The fitness ratio of the remaining population is $0.9983$ and it is greater than $y$. In this case, relying on it to predict that the candidate is fit would be incorrect because, unfortunately, the respondent is unfit. Despite this, the interview is still correct, as we have succeeded in reaching the required fitness ratio, which allows a margin of error.

    \item If the respondent is not of type \{2\}, then they will have answered 1 to one of the questions, since the only two candidate types that answer 0 to both \{1,2,4\} and \{1,3\} are \textit{null} and \{2\}. Therefore, we will have ruled out candidates of type \textit{null}, which are the only fit ones, and the fitness ratio of the remaining population will be sufficiently low ($0\leq x = 0$).
\end{itemize}

Regardless of who the respondent is, we can guarantee that we will arrive at a fitness ratio less than or equal to  $x=0$, or greater than or equal to $y = 0.9948$.

In order to conclude that the transformation is valid for our example, we check that the interview does not infer more private information than allowed. The ratio of the initial population with the private characteristic (i.e.\ those who answer 1 to private question) is  $|S| / (|U|*\Omega + \Omega^2 + |S|) = 0.0044$, and we need to ensure that it remains between $ a=(|S| - k) / (|U|*\Omega + \Omega^2 + |S| - k) = 0.0015$ and $b=1$.

First, suppose the answer given to question \{1, 2, 4\} is 0. The ratio of the remaining population (which consists of types 3, \textit{null}, \{1,3\}, and \{2\}) with the private characteristic is $2 / (\Omega^2 + \Omega + 2) = 0.0033$, and we remain within the interval $[a, b]$. Regarding the next question, i.e.\ \{1, 3\},

\begin{itemize}
    \item if the answer is 0, then the ratio of the remaining population (i.e.\ types \textit{null} and \{2\}) with the private characteristic is $1 / (\Omega^2 + 1) = 0.017 \in [0.0015, 1]$, and

    \item if the answer to question \{1, 3\} is 1, then the ratio of the remaining population (types 3 and \{1, 3\}) with the private characteristic is $1 / (\Omega + 1) = 0.04 \in [0.0015, 1]$.
\end{itemize}

On the other hand, suppose the answer given to the first question \{1, 2, 4\} is 1. The ratio of the remaining population (types 1, 2, 4, \{1,2,4\}) with the private characteristic is $1 / (3*\Omega + 1) = 0.0137 \in [0.0015, 1]$ and we will end the interview, as indicated by our strategy.

It can be proven in a similar manner that asking the question \{1, 3\} before \{1, 2, 4\} does not affect the result. This proof has been omitted due to is redundancy. Appendix~\ref{sec:appendix proof} shows, in the process of demonstrating that the reduction we have defined is correct, that the order in which we ask the questions does not matter.

As we have seen, by following the defined strategy it is not possible to infer more private information than allowed, and we can ensure that the transformation is correct for this particular instance of the set cover decision problem.

With this example, we can get a glimpse of the reasons why the proposed reduction works:
\begin{itemize}
    \item In most cases, it is not possible to ask the question \textit{private}, as it trivially infers private information. The only exception to this is when we have already ruled out all candidates that respond 0 to \textit{private}. This peculiar case is taken into account in Lemma~\ref{lem:1} introduced in Appendix~\ref{sec:demonstration}.

    \item We can only ask up to $k$ questions. Otherwise we would be ruling out more \textit{set} candidates than permitted, because each question rules out exactly 1 candidate of that type. If we rule out more than $k$ \textit{set} candidates, then the ratio of the remaining population with the private characteristic would be too low (since $a=(|S| - k) / (|U|*\Omega + \Omega^2 + |S| - k)$).\footnote{Note that, as long as the respondent keeps answering 0 and thus \textit{null} has not been ruled out, ruling out a single \textit{set} candidate always has to reduce the ratio of the population with the private characteristic (which initially is $|S| / (|U|*\Omega + \Omega^2 + |S|)$) regardless of whether, simultaneously, all \textit{element} candidates are ruled out. This is because the value of $\Omega$ has been chosen to be sufficiently high so that $1/|S| > (|U|*\Omega + |S|) / (|U|*\Omega + \Omega^2 + |S|)$ holds.}

    \item It is possible to rule out all the \textit{element} candidates with $k$ \textit{set} questions if and only if it is possible to cover all the elements in $U$ of the instance of \SCP\ with $k$ sets.

    \item It is mandatory to rule out all \textit{element} candidates if we want to satisfy the fitness requirements ($y=\Omega^2/(\Omega^2+|S|)$).
\end{itemize}

\subsection{Results of the analysis}
\label{sec:results}

After the informal presentation of results, we summarize the complexity analysis, which is fully developed in Appendix~\ref{sec:appendix proof}: we define a transformation algorithm of instances of the set cover decision problem \SCP\ to \DCDP\ and prove that it is polynomial-time (see Section~\ref{sec:transformation}); we show that if the result of an instance of \SCP\ is $\top$ then the result of its transformed instance will also be $\top$ (Appendix~\ref{sec:positive instances}); we  show that if the result of an instance of \SCP\ is $\bot$ then the result of its transformed instance will be $\bot$ (Appendix~\ref{sec:negative instances}); and finally we prove that \DCDP\ belongs to NP (Appendix~\ref{sec:appendix NP}).\footnote{Since solutions to this problem are {\it trees}, the inclusion in NP could be surprising at a first glance. The key will be that these trees will be sufficiently small, in particular polynomial-sized.} Therefore, we  conclude that the defined polynomial reduction is correct and that \DCDP\ is NP-complete.

Having established the NP-hardness of \DCDP, it follows that its optimization and more general variant, \GDCOP, is also NP-hard. As aforementioned, the NP-hardness is inherited from the more particular variants of a problem into the more general ones. Formally, we can demonstrate the NP-hardness of \GDCOP\ by using a very simple polynomial reduction.
Recall that instances of \DCDP\ and \GDCOP\ are the tuples of the form $(f, g, P, a, b, x, y)$ and $(f_g, g, P_g, k, x, y)$, respectively, where $f:(\mathit{Question}\rightarrow \mathit{Answer}) \rightharpoonup \{\top, \bot\}$, $f_g:(\mathit{Question}\rightarrow \mathit{Answer}) \rightharpoonup \{q\ |\ q \in \mathbb{Q}, 0\leq q\leq 1\}$, and $P_g\subseteq \mathit{Question}\times \mathit{Answer}\times \{a\ |\ a\in \mathbb{Q}, 0\leq a\leq 1\} \times \{b\ |\ b\in \mathbb{Q}, 0\leq b\leq 1\}$.

To transform in polynomial time instances of \DCDP\ into instances of \GDCOP, first we replace all occurrences of $\top$ and $\bot$ in the image of the fitness function $f$ by 1 and 0, respectively, to construct $f_g$. Remember that, while the possible outputs of $f$ are $\top$ (fit) and $\bot$ (unfit), the function $f_g$ returns rational numbers between $0$ and $1$. Then, in order to define $P_g$ we replace each private characteristic by a tuple with that characteristic, the answer 1 (as explained in Section~\ref{sec:formal definition}, in \DCDP\ we need to ensure that the ratio of the population that responds 1 to each private characteristic is maintained throughout the interview), and the bounds $a$ and $b$. That is, $P_g = \{ (p, 1, a, b)\ |\ p\in P\}$. For example, from $P=\{male, religious\}$ we would construct $P_g=\{(male, 1, a, b), (religious, 1, a, b)\}$. Finally, the question limit $k$ will be the total number of available questions, which is equivalent to not having a question limit.

It is clear that this transformation can be done in polynomial time with respect to the size of the input. Trivially, the resulting instance is equivalent to the original instance and thus this simple polynomial reduction from \DCDP\ is correct and \GDCOP\ is NP-hard.

\section{Practical resolution}
\label{sec:practical resolution}

The data classification problem with private characteristics is NP-hard and, thus, solving large instances in an exact way is not feasible. In order to deal with large instances, using heuristic methods is necessary. These methods do not guarantee finding a perfect solution, but can provide sufficiently good solutions within a reasonable time frame.

While the higher particularity of \DCDP\ makes it  more interesting for NP-hardness classification, \GDCOP\ is better suited for dealing with real situations and therefore it will be used in the experiments. 
The code used in the practical resolution 
is available in an online repository~\cite{pantoja2024tfm}.

Before delving into the algorithms, it is important to clarify the use of the term ``fitness''. So far, we have used ``fitness'' to refer to the aptitude of a candidate or group of candidates. When discussing genetic algorithms and other heuristic algorithms, ``fitness'' often refers to how good a solution of an instance of a problem is. To disambiguate this term, we are going to use \textit{goodness} to refer to how good an interview is. Recall that we are trying to find an interview that does not infer private information and that distinguishes the fit and unfit candidates as well as possible, making extreme (i.e.\ maximizing or minimizing) the fitness of the candidates in the leaves of the interview tree.

\subsection{Exact algorithm}
\label{sec:exact algorithm}

As we have discussed, solving large instances of the problem is unfeasible due to its proven complexity. To illustrate this, we have implemented in Python an algorithm that solves instances of the problem. This simple algorithm uses alpha-beta pruning~\cite{felstiner2019alpha} to decrease the solution space that needs to be evaluated.

To test the scalability of the algorithm with respect to the number of questions, we ran this algorithm on 5 randomly constructed instances of \GDCOP\ for each number of questions between 1 and 10 and we calculated the average run time. The randomly constructed instances of the problem have 500 different candidate types and up to 10 questions per interview ($k=10$). Table~\ref{tab:exact_n_questions} shows the results.

\begin{table}
    \centering
    \begin{tabular}{|c|c|} \hline
         \textbf{Number of questions}& \textbf{Run time (s)}\\ \hline
         1& 0.0009\\ \hline
         2& 0.049\\ \hline
         3& 5.8657\\ \hline
         4& 9.2263\\ \hline
         5& 70.0895\\ \hline
         6& 176.3103\\ \hline
         7& 241.6572\\ \hline
         8& 687.6727\\ \hline
         9& 1348.9556\\ \hline
         10&1880.9390\\ \hline
    \end{tabular}
    \caption{Run time of the exact algorithm depending on the number of questions}
    \label{tab:exact_n_questions}
\end{table}

These simple tests showcase the unfeasibility of the problem ---at this rate, with just 30 questions, we would require years of run time. For reference, we will approximately solve instances of \GDCOP\ of more than 3000 candidate types and more than 200 questions in the following sections. Thus, it would be completely impossible to deal with such instances using exact methods.

\subsection{Greedy algorithm}
\label{sec:greedy algorithm}

We implemented a simple greedy algorithm in Python to approximately solve instances of \GDCOP. This algorithm greedily chooses, at each node of the interactive interview tree under construction, the remaining question that maximizes the goodness of the interview without inferring private information. This choice is made without taking into consideration that, in general, the interview (i.e. tree) will continue growing in the same way in subsequent nodes.

In particular, in order to choose the question associated with each node in the tree, we consider it as if the interview would end with that question. We order the possible questions according to the goodness that the interview would have with them, and then we go through the ordered list of questions until we find, if there is one, a question that does not infer private information. If we find one then we choose that question at the current node and continue with the algorithm, choosing in the same way the rest of the questions in the tree. If we find none then we end that path of the interview. To speed up the algorithm, we keep a list of the questions that have not been asked in the path that is being explored. The following pseudocode illustrates this algorithm:

\begin{lstlisting}

Greedy(candidates, question_limit, questions)
   IF no questions left OR question limit reached
      RETURN Empty interview

   sort questions by the population fitness after each question

   FOR each question left
      FOR each possible answer to the question
         IF answer does not infer private information
            exit the for loops

   IF every question infers private information
      RETURN Empty interview

   children = Empty map
   FOR each possible answer to question
      matching_candidates = those who respond answer to question
      children[answer] = Greedy(matching_candidates,
         question_limit-1, questions - question)

   RETURN Interview(question, children)

\end{lstlisting}

\subsection{Genetic algorithms}
\label{sec:genetic algorithm}

A genetic algorithm \cite{katoch2021review,alhijawi2024genetic} is a type of heuristic optimization algorithm inspired by biological evolution and genetics that consists in evolving a population of individuals, where each individual encodes a possible solution to the problem, through crossovers and mutations. Briefly, the basic operation of a genetic algorithm is as follows:

\begin{enumerate}
    \item First, the initial population is created.
    \item Individuals from the population are crossed in pairs, generating new individuals with traits from both parents.
    \item Each individual has a small probability of mutating, slightly modifying some of its traits.
    \item Based on the goodness of the individuals, those that will be part of the population for the next iteration are selected.
    \item This process is continued iteratively, returning to point 2.
\end{enumerate}

After a given number of iterations, the algorithm will terminate, returning the solution coded by the best individual.

\begin{figure}
    \includegraphics[width=0.8\linewidth]{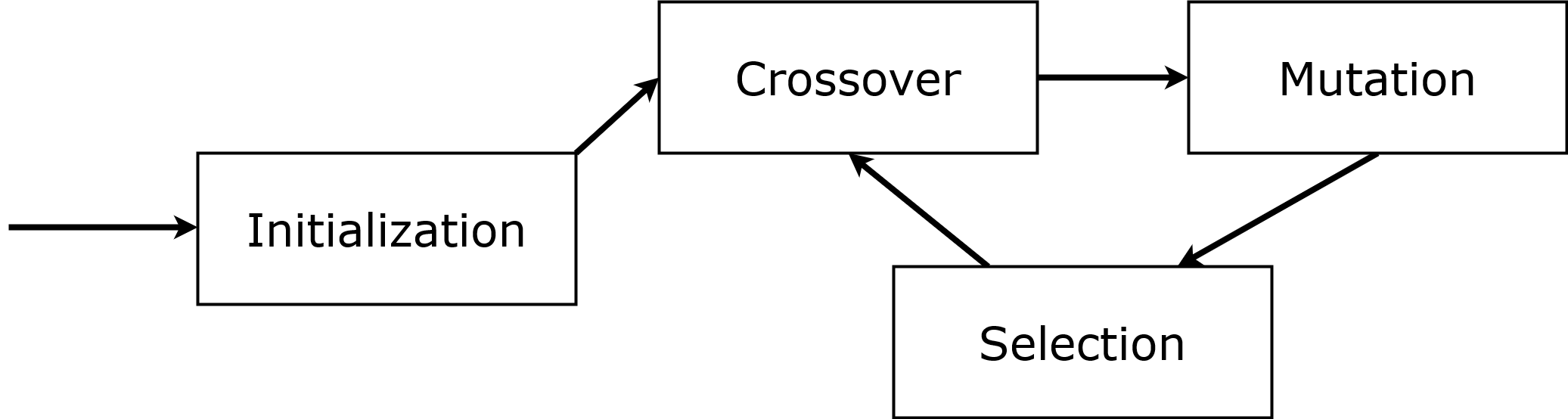}
    \caption{Schematic of a basic genetic algorithm}
    \label{fig:gen alg}
\end{figure}

In our Python implementation of the genetic algorithm, we encode the solutions (i.e. interactive interviews) of the instance of \GDCOP\ being studied as trees where the values of the nodes are the questions. At no time  any interview in the population will be able to infer private information.

Using a genetic algorithm to solve our data classification problem is not trivial. We have to take into account that, generally, the majority of arbitrarily constructed interviews infer private information in at least one of the paths of the tree. Additionally, interviews are only as good as their worst path. An interview that obtains good results for the majority of the candidates but infers private information about a few candidates is incorrect. This makes it difficult for the genetic algorithm to eventually generate correct interviews when the initial population consists of incorrect interviews.

Because of this, we are going to guarantee that the initialization, crossover and mutation methods do not generate incorrect interviews. If at any point we add to an interview a question that infers private information, we will try to find another. Due to the size of the interviews, this process can affect the performance of the genetic algorithm. We are going to limit the number of times that we will be allowed to try to find a correct question to 100 attempts per initialization or mutation of an interview. If we run out of tries or there are no more questions left, then the path of the interview will end.

The initialization of the population consists in creating question and answer trees randomly, but ensuring that the generated interviews are not incorrect.

To generate new individuals by crossing two individuals, we create a new tree with a question at the root node that is not present in the interviews of the parents. For each possible answer to that question, its resulting subtree is one of the parents chosen randomly. To ensure the correctness of the generated tree, the paths that infer private information are pruned.

In each iteration and for each interview, the mutation function has a certain probability of replacing the subtree of one of the nodes of the interview tree (located at a random depth) with a random subtree that guarantees the correctness of the interview.

As selection function we apply the tournament method. That is, we pair individuals randomly and keep the best individual from each pair for the next iteration of the algorithm.

The goodness function (the fitness function for the individuals of the genetic algorithm) assigns values to the interviews based on the probability of correctly predicting a candidate's fitness in the worst case. To accomplish this, it returns the ratio of fit candidates or the ratio of unfit candidates, whichever is higher, in the leaf of the tree whose ratios of fit and unfit candidates are closest.

In addition to this basic genetic algorithm, we have implemented a genetic algorithm reinforced with the greedy algorithm presented earlier (see Section~\ref{sec:greedy algorithm}). This algorithm differs from the previous one in that, at initialization, one interview is generated directly by the greedy algorithm and another is a mutation of the greedy solution. The rest of the population is initialized in the same way as with the basic genetic algorithm. In addition, half of the mutations replace a subtree of the interview with the solution that we would obtain with the greedy algorithm from that point onward, rather than with a randomly generated subtree.

\subsubsection{Parameter tuning}
\label{sec:parameter tuning}

In this section, we explain how we tuned the parameters of the genetic algorithms. For this purpose, we generated random instances of \GDCOP. These constructed instances have up to 5 answers per question, 100 candidates per candidate type, and do not have many private characteristics (in particular, 4-9), as it is the most common scenario in real interviews. Moreover, instances with many private characteristics are not very useful for comparison purposes because they strongly limit the size that the interviews can have without inferring private characteristics.

We ran the basic genetic algorithm on 10 different randomly constructed instances of \GDCOP\ to choose the size of the population. We used a mutation ratio of 0.1 and we executed the genetic algorithm for 200 iterations. Table~\ref{tab:instances_population} shows the sizes of the instances and Table~\ref{tab:size_population} the results obtained. The results illustrate that using 20 individuals is clearly better than using only a population size of 10 elements. However, increasing the population size up to 30 or 40 individuals does not improve the quality of the results. In fact, the statistical tests we applied show that there is indeed a statistically significant difference between 10 and 20 individuals, but not between 20 and 30 or between 20 and 40. Thus, we decided to use a population size of 20 for the rest of our experiments.

\begin{table}
    \centering
    \begin{tabular}{|c|c|c|c|c|} \hline
         \textbf{Instance}&  \textbf{Candidate types}&  \textbf{Char.}&  \textbf{Private char.}& \textbf{Question limit}\\ \hline
         1&  3097&  225&  7& 10\\ \hline
         2&  3303&  224&  6& 11\\ \hline
         3&  2165&  227&  8& 12\\ \hline
         4&  2254&  180&  6& 13\\ \hline
         5&  2602&  195&  6& 14\\ \hline
         6&  3536&  152&  9& 12\\ \hline
         7&  3954&  239&  6& 13\\ \hline
         8&  2158&  235&  8& 12\\ \hline
         9&  3295&  235&  9& 14\\ \hline
         10&  3963&  262&  6& 15\\ \hline
    \end{tabular}
    \caption{Overview of the instances used to tune the size of the population}
    \label{tab:instances_population}
\end{table}

\begin{table}
    \centering
    \begin{tabular}{|c|c|c|c|c|} \hline
         \textbf{Instance} &  \textbf{Size 10}&  \textbf{Size 20}&  \textbf{Size 30}& \textbf{Size 40}\\ \hline
         1 &  0.8463&  0.8827&  0.8669& 0.9102\\ \hline
         2 &  0.9380&  0.9050&  0.9079& 0.9334\\ \hline
         3 &  0.8680&  0.8910&  0.9051& 0.8351\\ \hline
         4 &  0.9218&  0.9845&  0.9555& 0.9284\\ \hline
         5 &  0.8851&  0.9690&  0.9317& 0.9534\\ \hline
         6 &  0.7806&  0.8209&  0.9669& 0.8783\\ \hline
         7 &  0.8628&  0.8660&  0.8138& 0.9023\\ \hline
         8 &  0.8582&  0.9021&  0.8965& 0.8889\\ \hline
         9 &  0.8043&  0.8714&  0.9507& 0.8767\\ \hline
         10 &  0.8268&  0.8716&  0.8394& 0.9071\\ \hline
         Average &  0.8592&  0.8964&  0.9035& 0.9014\\ \hline
    \end{tabular}
    \caption{Performance of the genetic algorithm depending on the size of the population}
    \label{tab:size_population}
\end{table}

\begin{figure}
\begin{tikzpicture}
\begin{axis}[
    xlabel={Number of iterations},
    ylabel={Average best solution obtained},
    xmin=0, xmax=500,
    ymin=0.6, ymax=1,
    xtick={0,100,200,300,400,500},
    ytick={0.6,0.7,0.8,0.9,1},
    legend pos=north west,
    ymajorgrids=true,
    grid style=dashed,
]
\addplot[
    color=blue,
    mark=square,
    ]
    coordinates {
    (1,0.6199363152)
    (10,0.7687150891)
    (20,0.7856164185)
    (30,0.8143750065)
    (40,0.8205554515)
    (50,0.8364753386)
    (100,0.8594203562)
    (150,0.8631523857)
    (200,0.8666845006)
    (250,0.8818647368)
    (300,0.8818647368)
    (350,0.8818647368)
    (400,0.8947088845)
    (450,0.8947088845)
    (500,0.8951256987)
    };
\end{axis}
\end{tikzpicture}
\caption{Performance of the genetic algorithm depending on the number of iterations}\label{graph:iterations}
\end{figure}
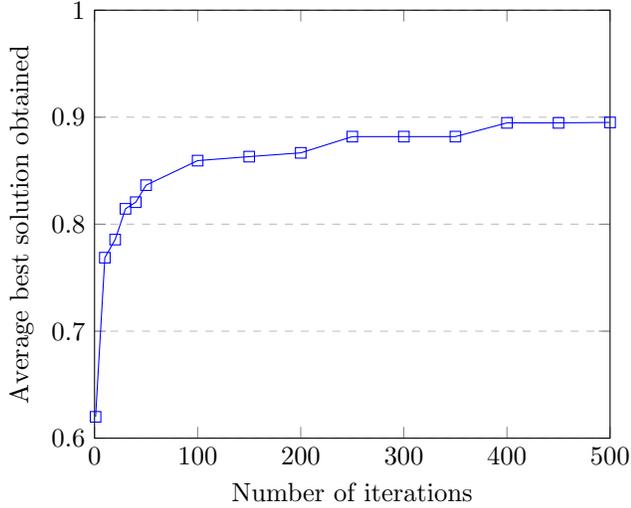

Similarly, to choose the number of iterations, we ran 10 random instances with a mutation ratio of 0.1 and a population size of 20 and periodically measured the best solution obtained. Table~\ref{tab:instances_iterations} shows the sizes of the instances and Figure~\ref{graph:iterations} illustrates the performance of the algorithms depending on the number of iterations. As it can be seen, the degree of improvement decreases significantly as the iterations progress, so that after 400 iterations there is no substantial difference with further iterations. Thus, we chose to set the maximum number of iterations of our genetic algorithms at 400.

\begin{table}
    \centering
    \begin{tabular}{|c|c|c|c|c|} \hline
         \textbf{Instance}&  \textbf{Candidate types}&  \textbf{Char.}&  \textbf{Private char.}& \textbf{Question limit}\\ \hline
         1&  2823&  168&  9& 14\\ \hline
         2&  3962&  251&  4& 14\\ \hline
         3&  3199&  298&  6& 14\\ \hline
         4&  2275&  180&  7& 15\\ \hline
         5&  3702&  261&  9& 11\\ \hline
         6&  3290&  181&  4& 14\\ \hline
         7&  3552&  259&  8& 15\\ \hline
         8&  2684&  243&  8& 15\\ \hline
         9&  3417&  210&  4& 11\\ \hline
        10&  3492&  200&  5& 11\\ \hline
    \end{tabular}
    \caption{Overview of the instances used to tune the number of iterations}
    \label{tab:instances_iterations}
\end{table}

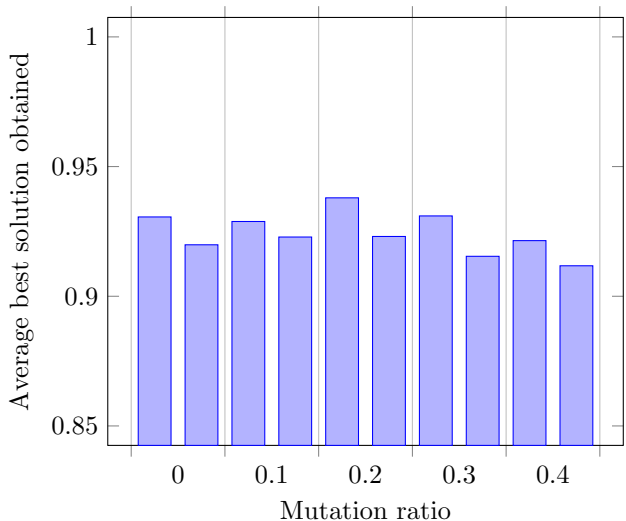
\begin{figure}
\begin{tikzpicture}
\begin{axis}[
	x tick label style={
		/pgf/number format/1000 sep=},
    xlabel={Mutation ratio},
	ylabel={Average best solution obtained},
	enlargelimits=0.05,
	legend style={at={(0.5,-0.1)},
	anchor=north,legend columns=-1},
	ybar interval=0.7,
    xtick={0,0.1,0.2,0.3,0.4,0.5},
    ytick={0.85,0.9,0.95,1},
    xmin=0, xmax=0.5,
    ymin=0.85, ymax=1,
]
\addplot
    coordinates {
    (0,0.9305784997)
    (0.05,0.9198425599)
    (0.1,0.9288262612)
    (0.15,0.9228357304)
    (0.2,0.9379467379)
    (0.25,0.9230511742)
    (0.3,0.9309718323)
    (0.35,0.915405341)
    (0.4,0.9214466807)
    (0.45,0.9117507045)
    (0.5,0.9317448356)
    };
\end{axis}
\end{tikzpicture}
\caption{Average performance in terms of the mutation ratio}\label{graph:mutation}
\end{figure}

Finally, to choose the mutation ratio of the genetic algorithm, we ran 10 random instances of \GDCOP\ with a population size of 20 individuals and for 400 iterations with mutation ratios between 0 and 0.5. Table~\ref{tab:instances_mutation} shows the sizes of the instances and Figure~\ref{graph:mutation} illustrates the results obtained in the runs. It can be seen that the best results are obtained using 0.2 as mutation ratio. However, different mutation rates seem to have a
very reduced
effect on the performance of the algorithm. The main purpose of the mutation of individuals in genetic algorithms is to facilitate the exploration of the solution space. Due to its randomness, the crossover algorithm already serves this purpose. Additionally, an interview is only as good as its worst path, and, due to the size of the interview trees, it is likely that the worst path of an interview remains unchanged after a mutation.
Thus, the results shown in Figure~\ref{graph:mutation}
do not show a clear tendency, but the best results are obtained using ratio 0.2.

\begin{table}
    \centering
    \begin{tabular}{|c|c|c|c|c|} \hline
         \textbf{Instance}&  \textbf{Candidate types}&  \textbf{Char.}&  \textbf{Private char.}& \textbf{Question limit}\\ \hline
         1&  2470&  181&  9& 15\\ \hline
         2&  2251&  217&  4& 15\\ \hline
         3&  3999&  301&  8& 11\\ \hline
         4&  2287&  279&  4& 15\\ \hline
         5&  2852&  301&  6& 15\\ \hline
         6&  2927&  187&  4& 11\\ \hline
         7&  2341&  252&  5& 12\\ \hline
         8&  2796&  270&  6& 14\\ \hline
         9&  2838&  174&  9& 13\\ \hline
         10&  2434&  153&  9& 13\\ \hline
    \end{tabular}
    \caption{Overview of the instances used to tune the mutation ratio}
    \label{tab:instances_mutation}
\end{table}

Given the results obtained, we decided that the parameters to be used in the rest of the practical resolution are 400 iterations, a population size of 20 and a mutation rate of 0.2.

\subsection{Comparison of the performance of the algorithms}
\label{sec:comparison algorithms}

This section aims to show and compare the performance of the implemented heuristic algorithms. We tested the performance of the greedy algorithm, the basic genetic algorithm and reinforced genetic algorithm ---recall that the latter is the genetic algorithm including the greedy solution as well as a mutation of it in its initial population--- with 50 randomly constructed instances of \GDCOP, using the chosen parameters in the genetic algorithms. We ran each genetic algorithm 3 times per instance, keeping the best result obtained. Tables \ref{tab:instances_comparisons} and \ref{tab:comparison} show the size of the instances and the results obtained, respectively.

\begin{table}
	\centering
    {\scriptsize
	\begin{tabular}{|c|c|c|c|c|} \hline
     	\textbf{Instance}&  \textbf{Candidate types}&  \textbf{Char.}&  \textbf{Private char.}& \textbf{Question limit}\\ \hline
     	1&  2074&  295&  4& 15\\ \hline
     	2&  2748&  279&  7& 15\\ \hline
     	3&  2247&  240&  6& 14\\ \hline
     	4&  2144&  184&  7& 11\\ \hline
     	5&  3248&  267&  6& 11\\ \hline
     	6&  2084&  157&  7& 10\\ \hline
     	7&  2058&  284&  7& 15\\ \hline
     	8&  2722&  189&  8& 12\\ \hline
     	9&  3614&  281&  7& 14\\ \hline
     	10&  2349&  257&  6& 12\\ \hline
 11& 3096& 191& 5&10\\\hline
 12& 2831& 283& 7&10\\\hline
 13& 2086& 221& 6&14\\ \hline
 14& 3852& 159& 9&13\\ \hline
 15& 2622& 229& 6&14\\ \hline
 16& 2164& 199& 8&10\\ \hline
 17& 2089& 274& 8&11\\ \hline
 18& 2406& 152& 9&11\\ \hline
 19& 2916& 254& 9&14\\\hline
 20& 2839& 184& 9&12\\\hline
 21& 2551& 231& 6&14\\\hline
 22& 2272& 161& 4&12\\\hline
 23& 3808& 240& 8&13\\\hline
 24& 3208& 282& 5&11\\\hline
 25& 3313& 230& 7&11\\\hline
 26& 3487& 178& 4&15\\\hline
 27& 2056& 241& 7&14\\\hline
 28& 2608& 208& 9&10\\\hline
 29& 3559& 166& 6&15\\\hline
 30& 2406& 281& 4&14\\\hline
 31& 2169& 156& 4&14\\\hline
 32& 3257& 221& 7&12\\\hline
 33& 3065& 223& 8&13\\\hline
 34& 3025& 301& 5&10\\\hline
 35& 2593& 235& 8&15\\\hline
 36& 3794& 298& 9&12\\\hline
 37& 2471& 152& 4&15\\\hline
 38& 3714& 281& 9&15\\\hline
 39& 2924& 193& 6&10\\\hline
 40& 2596& 286& 5&15\\\hline
 41& 2841& 214& 9&14\\\hline
 42& 3601& 211& 9&11\\\hline
 43& 2630& 278& 8&13\\\hline
 44& 2204& 178& 4&14\\\hline
 45& 2851& 290& 9&14\\\hline
 46& 3218& 221& 8&13\\\hline
 47& 2076& 302& 9&13\\\hline
 48& 2337& 231& 8&13\\\hline
 49& 3709& 172& 6&14\\\hline
 50& 2936& 289& 6&15\\\hline
	\end{tabular}
}
	\caption{Overview of the instances used to compare the heuristic algorithms}
	\label{tab:instances_comparisons}
\end{table}

\begin{table}
	\centering
{\scriptsize
	\begin{tabular}{|c|c|c|c|} \hline
     	\textbf{Instance}&  \textbf{Greedy}&  \textbf{Basic genetic}& \textbf{Reinforced genetic}\\ \hline
     	1&  0.7285&  1& 1\\ \hline
     	2&  0.522&  0.9573& 1\\ \hline
     	3&  0.7145&  1& 1\\ \hline
     	4&  0.5098&  1& 1\\ \hline
     	5&  0.6053&  0.9637& 1\\ \hline
     	6&  0.6368&  1& 1\\ \hline
     	7&  0.5015&  0.9396& 0.9562\\ \hline
     	8&  0.5314&  0.912& 1\\ \hline
 9& 0.5257& 0.9137&0.9188\\ \hline
 10& 0.5163& 0.9632&0.9266\\ \hline
 11& 0.5471& 0.9455&1\\ \hline
 12& 0.5147& 0.9343&0.9537\\ \hline
 13& 0.5724& 1&0.9917\\ \hline
 14& 0.5009& 0.8708&0.9462\\ \hline
 15& 0.609& 1&1\\ \hline
 16& 0.5152& 0.8532&0.848\\ \hline
 17& 0.5095& 0.9347&0.9434\\ \hline
 18& 0.623& 0.9495&0.9021\\ \hline
     	19&  0.5127&  0.9105& 0.9095\\ \hline
 20& 0.5071& 0.9613&0.9039\\ \hline
 21& 0.6304& 0.9456&0.9362\\ \hline
 22& 0.6152& 1&1\\ \hline
 23& 0.5534& 0.9096&1\\ \hline
 24& 0.6173& 0.9074&0.9416\\ \hline
 25& 0.586& 0.9313&1\\ \hline
 26& 0.5014& 1&0.993\\ \hline
 27& 0.6054& 0.9361&1\\ \hline
 28& 0.5344& 0.9731&1\\ \hline
 29& 0.5403& 0.9182&1\\ \hline
 30& 0.5066& 0.9432&0.9212\\\hline
 31& 0.503& 0.99&0.9684\\ \hline
 32& 0.5011& 1&1\\ \hline
 33& 0.5417& 0.9649&0.9067\\ \hline
 34& 0.5151& 0.8824&0.9184\\ \hline
 35& 0.5091& 0.9223&0.9255\\ \hline
 36& 0.5112& 0.9012&0.8874\\ \hline
 37& 0.5088& 0.906&0.9255\\ \hline
 38& 0.5008& 0.8821&0.9114\\ \hline
 39& 0.5017& 0.9716&0.8871\\ \hline
 40& 0.5089& 0.9403&0.9895\\\hline
 41& 0.6156& 0.9695&1\\ \hline
 42& 0.5025& 0.9445&0.8876\\ \hline
 43& 0.5602& 0.9057&0.8735\\ \hline
 44& 0.5107& 1&1\\ \hline
 45& 0.5089& 0.9925&0.9632\\ \hline
 46& 0.5726& 0.8957&0.9465\\ \hline
 47& 0.5162& 0.9308&0.9416\\ \hline
 48& 0.5075& 0.8919&0.9606\\ \hline
 49& 0.5554& 0.9289&1\\ \hline
 50& 0.6585& 0.9721&1\\\hline
 Average goodness& 0.5501& 0.9453&0.9577\\\hline
	\end{tabular}
}
	\caption{Comparison of the performance of the greedy, basic genetic and reinforced genetic algorithms}
	\label{tab:comparison}
\end{table}

It is noticeable that the greedy algorithm performs significantly worse than the genetic algorithms. Note that even the worst correct interview for any given instance of the problem has a goodness of at least 0.5. This can only occur in instances where exactly 50\% of candidates are fit, and when the interview does not discern fit from unfit candidates in any way in the worst case (there exists a leaf of the interview tree where 50\% of candidates are fit). Additionally, there are instances where it is not possible to completely discern fit and unfit individuals without inferring private information, and thus there will not exist any interview for them with a goodness of 1. Therefore, the difference from an average goodness of 0.5501 to 0.9453 or 0.9577 is more significant than it may appear at first glance. However, the difference between the basic genetic algorithm and the reinforced genetic algorithm is not as clear. In fact, a statistical test is necessary to assess whether there is a statistically significant difference between both methods.

In order to do that, we used the STAC tool \cite{rodriguez2015stac} to compare the results obtained. STAC is a web platform and a Python library for comparing algorithms using statistical tests.
We performed the comparison of the genetic algorithms with the binomial sign test \cite{conover1999signtest}. This non-parametric (it is not assumed that the data follows any specific distribution) two-group hypothesis test compares paired observations, which in our case are the goodness of the best interviews found by the genetic algorithms for each instance, and does not assume that the differences between the observations are symmetrical with respect to the median. We have used the binomial sign test to detect the presence of significant differences in the performance of the genetic algorithms.

\begin{table}
	\centering
	\begin{tabular}{|c|c|c|} \hline
     	\multicolumn{3}{|c|}{\textbf{Binomial sign test ($\alpha=0.1$)}}\\ \hline
     	\textbf{Statistic}&  \textbf{\textit{p} value} & \textbf{Result}\\ \hline
     	26&  0.088& $H_0$ is rejected\\ \hline
	\end{tabular}
	\caption{Results of the binomial sign test}
	\label{tab:prac_res_1}
\end{table}

Table \ref{tab:prac_res_1} shows the results obtained. The \textit{statistic} is a measure calculated from the results of the algorithms that is used to determine whether the null hypothesis $H_0$ (i.e.\ the absence of statistical difference in the comparison) is plausible. The significance level ($\alpha$) is the probability of rejecting $H_0$ when it is true, i.e. the probability of false positives. The value \textit{p} indicates the probability of observing the results obtained by the algorithms assuming that $H_0$ were true. When this value is less than $\alpha$, the test is considered statistically significant and $H_0$ is rejected.

Finally, to ensure that the reinforced genetic algorithm returns proper results for the small instances of the problem that the exact algorithm is able to solve, we have compared their results with 10 randomly generated instances of 500 candidate types and 10 questions, obtaining the same results.

With these tests, we show that the reinforced genetic algorithm performs significantly better than the basic genetic algorithm and the greedy algorithm separately. The greedy mutations and the completion of solutions with greedy subtrees help to steer the genetic algorithm when it cannot find a way to improve its population interviews.
All in all, we can conclude that the reinforced genetic algorithm clearly outperforms the other two methods.

\section{Conclusions and future work}
\label{sec:conclusions}

In this paper we have formally introduced the data classification problem with private characteristics as well as an optimization variant.
We analyzed the computational complexity of the problem, and for this purpose a polynomial reduction from the NP-complete set cover decision problem was developed. Beyond the relevance of the classification of the proposed problem itself, this can help to classify the complexities of other future (apparently or subtly) related problems.

Due to the computational complexity of the problem, optimally solving large instances is not feasible, so we implemented heuristic algorithms that seek suboptimal solutions in a reasonable time. We implemented a greedy algorithm, a genetic algorithm and a genetic algorithm reinforced with the greedy algorithm, and compared their performances with a concrete benchmark.
We observed that the genetic algorithm reinforced with the greedy algorithm performs better than the others, with a statistically significant difference.

Our proof that the problem is NP-complete indicates that there are no simple strategies to ensure that the best possible tradeoffs between privacy protection and classification accuracy are obtained. Moreover, our experimental results corroborate that using simple strategies (such as our greedy algorithm) yields rather poor results. Thus, the solution is to look for heuristic methods that can be adapted to each specific case, exploiting different characteristics in each situation. In this context, the use of genetic algorithms is a very good point of balance, since they allow us to have a general method that is able to adapt to each specific case study.
Thus, our recommendation is that whenever we are faced with a problem where it is necessary to maintain the balance between privacy protection and classification accuracy, our genetic algorithms should be used directly.

The problem introduced in this paper, that is, the data classification problem with private characteristics, is of great significance in the current context due to its numerous applications. It facilitates the development of more efficient and ethical solutions in areas such as the design of interviews granting an opportunity (e.g. interviews determining access to a job, college, benefit,  adoption, etc.), user data gathering, certification processes guaranteeing privacy, etc.

It is worth noting that one could consider two different kinds of data to be protected in the context of our problem:
(a) the statistical information about the population; and (b) the information about the interviewed person in particular. Our problem deals only with protecting (b) because (a) is assumed to be {\it common knowledge}.
Let us recall that the dataset of statistical data of the population denotes the proportion of people with each specific combination of features (e.g. having features A, B, and C but not having D, E and F) who fulfill the target property we want to learn about ---as well as the proportion of them who fulfill the ``forbidden'' property we do not want to learn about. As mentioned before, these data (or, at least, its general trends) are either well known by experts in the corresponding fields or are publicly available via polls. For instance, the statistical relation between the level of education and the job performance is known in the literature; the usual diet restrictions of followers of some religions are well-known to any educated person; the general differences in consumption habits by gender, age, and wealth are known by marketing experts as well as anyone who wishes to read about them in the literature; etc. We cannot apply any data privacy protection to these data because it is common knowledge and it does not generally {\it belong} to the institution developing the adaptive interview. The institution could add some extra information about its own employees or former job applicants, but as long as the trends within the general population generally apply to the expected job applicants of the institution, the correlations leading to the potential uncertainty reduction about the target property or about the forbidden property are both well known to anyone. That is, the statistical dataset cannot be changed for the sake of privacy protection because virtually everybody knows it already ---or is just a single internet search away from knowing it.

Regarding (b), i.e. the privacy about the interviewed person, this is precisely the goal of the problem and the methods developed in the paper: highly reducing the uncertainty about the feature concerning the legitimate goal of the interview (e.g. the applicant aptitude to the job) while reducing as lowly as possible the uncertainty about the feature deemed private (e.g. the sexual orientation). After the designed adaptive interview is applied to the applicant, the partial revelation about the private feature due to the provided answers will stay under the established threshold just by its construction. As long as this threshold was set to be the desired one before the interview construction, there is no need to give any special treatment to the answers provided by the applicant after the interview: by definition, they respect the established threshold. Hence, no further post-processing action is needed to respect the desired level of privacy.

The possible future work of this study is vast. More problem variants should be defined and analyzed, which could fit better other possible real applications (e.g. considering the particular case where the interview is not adaptive).

Another possible future line of work is to develop and extend the practical resolution. More heuristic algorithms, such as the particle swarm optimization algorithm~\cite{wang2018particle}, can be implemented to solve this problem.
In this respect, it would be interesting to also explore the use of graph-based multi-dimensional and constrained machine learning approaches (see, e.g.~\cite{tutsoy2023graph}).
Note that while it may be complex to provide a generic method to solve our problem using this graph-based approach, specific instances of our problem could perhaps benefit from such an approach.

More heuristic algorithms, such as the particle swarm optimization algorithm, can be implemented to solve this problem. In addition, we could test the algorithms implemented with more intentionally constructed instances. The difficulty of an instance does not depend solely on its size; in many problems, such as 3-SAT, the difficulty of their instances is affected by specific patterns of their parameters, and investigating the effect of the relations between the candidates and questions of the instances of our problem on the performance of heuristic algorithms may be of interest.

\bigskip
\noindent\textbf{Code availability:}
Algorithms, data and data generation are available at {\scriptsize
\url{https://github.com/davidpantojasanchez/Data-classification-problem-with-private-characteristics.git}}.

\bigskip
\noindent\textbf{Author contributions:}
{\bf David Pantoja:} Writing– review \& editing, Writing– original draft,
 Visualization, Validation, Software, Resources, Methodology, Investigation, Formal analysis, Data curation, Conceptualization. {\bf Ismael Rodríguez:} Writing– review \& editing, Writing– original draft, Validation, Supervision, Resources, Methodology, Investigation, Formal analysis, Conceptualization. {\bf Fernando Rubio:} Writing– original draft, Validation, Supervision, Resources, Project administration, Methodology,
 Investigation, Funding acquisition, Formal analysis, Conceptualization,
 Writing– review \& editing. {\bf Clara Segura:} Writing– review \& editing, Writing– original draft, Validation, Supervision, Resources, Methodology, Investigation, Formal analysis, Conceptualization.


%

\section*{Conflict of interest}
The authors declare that they have no conflict of interest.

\bibliographystyle{spmpsci}      
\bibliography{biblio}   

%
%

\appendix
\section{Proof of the NP-completeness of the problem}
\label{sec:appendix proof}
In this appendix we provide the formal proof of the NP-completeness of \DCDP. First, we prove its NP-hardness by proving the correctness of the transformation that was sketched in Section~\ref{sec:complexity analysis}. Then, we prove that the problem belongs to NP, so that it is not only NP-hard but also NP-complete.

\subsection{Proof of the correctness of the transformation}
\label{sec:demonstration}

In this section we demonstrate the validity of the reduction of \SCP\ to \DCDP\ defined in Section~\ref{sec:transformation} and illustrated in the example given in Section~\ref{sec:transformation example}. Recall that an instance of \SCP\ is a tuple $(U, S, k)$, an instance of \DCDP\ is a tuple $(f, g, P, a, b, x, y)$, and we are assuming $|S|\neq k$.

We structure the demonstration as follows:

\begin{itemize}
    \item We define and verify two useful lemmas.

    \item We show that if the result of any instance of \SCP\ is $\top$ (positive instance) then, in the transformed instance of \DCDP, the strategy we defined will lead to a correct interview and the result will be $\top$.

    \item We show that if the result of any instance of \SCP\ is $\bot$ (negative instance) then
    it is not possible
    to produce an interview that meets the requirements in the transformed instance of \DCDP.
\end{itemize}

\begin{lemma}\label{lem:1}
If there exists a correct interview that includes the private question, there must exist a correct interview that is identical to the original except that it does not contain the private question.
\begin{proof}
If when asking the question called \textit{private} there are no candidates who would answer 0 or candidates who would answer 1, then the answer to the question would not narrow down the remaining population, as all remaining candidates would respond to question \textit{private} in the same way. Therefore, the question would be redundant and could be eliminated.

On the other hand, let us assume that when asking question \textit{private} there are candidates who would answer 0 and candidates who would answer 1 to that question. We know that $a>0$ because we are assuming that $|S|\neq k$, the definition of the problem ensures that $k\in \mathbb{N}$ and $k\leq |S|$, and $a=(|S| - k) / (|U|*\Omega + \Omega^2 + |S| - k)$. Since $a>0$ and there is at least one candidate who answers 0 and one who answers 1 to the private question, if any interview includes the private question then, after that question is responded, we cannot ensure that the ratio of candidates that who would answer 1 to the private question is greater than or equal to $a$; this is not true when the respondent answers 0. Therefore, all interviews that include the private question infer more private information than allowed in at least one of their paths and thus are incorrect.

Having proven that (a) no interview that asks question \textit{private} when there are still candidates remaining that would answer 0 and that would answer 1 is correct; and (b) that any interview that asks \textit{private} when the only remaining candidates would answer either 0 or 1 can be transformed into an equivalent one without question \textit{private}, we can ensure the result.
\end{proof}
\end{lemma}

\begin{lemma}\label{lem:2}
An interview of up to $k$ questions without  question \textit{private} asked to a \textit{null} candidate rules out all \textit{element} candidates if and only if the questions in the interview correspond to the sets in the solution of the original instance of \SCP.
\begin{proof}
\textit{Element} candidates answer 1 to the questions associated to the sets in $S$ to which their associated element belongs, and 0 to other questions. Thus, each time a \textit{null} respondent answers the question associated with a set $C\in S$, we rule out all \textit{element} candidate types whose associated element belongs to $C$, since we know that these candidates would answer 1 to this question and that \textit{null} candidates answer 0 to all questions; and we do not rule out any other \textit{element} candidate. Furthermore, as the interview cannot include the question \textit{private}, all questions in a valid interview must correspond to sets. Therefore, for an interview of up to $k$ questions to rule out all \textit{element} candidates, the union of the sets associated with the questions in the interview must be equal to $U$, and the result is true.
\end{proof}
\end{lemma}

\subsubsection{Analysis for positive instances of the set cover decision problem}
\label{sec:positive instances}

For positive instances of \SCP, there must be at least one set of sets of elements $R\in S$, $|R|\leq k$, such that $\cup_{A\in R}A=U$. As we saw in the example given in Section~\ref{sec:transformation example}, if the respondent is fit then they can only be of the candidate type \textit{null}. The interview resulting from the strategy we have defined will be a sequence of questions associated with the sets that belong to $R$.

 We start by studying the positive instances when the respondent is fit. Since the \textit{null} candidates answer 0 to all the questions and, according to our strategy, we do not finish the interview until we have asked all the questions in $R$ or have received a 1 as an answer, we can be sure that we will ask all the questions in $R$. Being in a positive instance, Lemma~\ref{lem:2} guarantees that the interview will rule out all \textit{element} candidates.

Barring \textit{private}, each question rules out exactly one of the \textit{set} candidate types, in particular the type that has the same associated set. For example, question \{2\} will rule out candidates of type $\{2\}$. Thus, we will rule out up to $k$ \textit{set} candidate types in total.

The ratio of candidates with the private characteristic in the initial population is $|S| / (|U|*\Omega + \Omega^2 + |S|)$. After the interview, we will have ruled out all the \textit{element} candidates and $1\leq k'\leq k$ \textit{set} candidate types (recall that we are assuming that $U\neq \emptyset$ because, as seen in Section~\ref{sec:transformation}, we separated that trivial case in the transformation, and therefore $0<k'$) and the ratio will be $(|S| - k') / (\Omega^2 + |S| - k')$. This ratio is between $a = (|S| - k) / (|U|*\Omega + \Omega^2 + |S|)$ and $b=1$:

\begin{equation}
k, k', |S|, |U| \in \mathbb{N};\ \ \ k' \leq k < |S|            
\end{equation}

\begin{equation} 
(|S| - k) / (|U|*\Omega + \Omega^2 + |S|) \leq (|S| - k) / (\Omega^2 + |S| - k')
\end{equation}

\begin{equation} 
(|S| - k) / (\Omega^2 + |S| - k') \leq (|S| - k') / (\Omega^2 + |S| - k')
\end{equation}

\begin{equation}
(|S| - k) / (|U|*\Omega + \Omega^2 + |S|) \leq (|S| - k') / (\Omega^2 + |S| - k') \leq 1
\end{equation}

We know (1) to be true, (2) and (3) hold if (1) is true, and (4) holds if (2) and (3) are true.

The ratio of fit candidates in the initial population is $\Omega^2 / (|U|*\Omega + \Omega^2 + |S|)$. After the interview, the ratio will be $\Omega^2 / (\Omega^2 + |S| - k)$, which is greater than or equal to $y=\Omega^2 / (\Omega^2 + |S|)$. Because of this, the transformation is correct for positive instances when the respondent is fit.

Next we look at positive instances when the respondent is not fit. Candidate types that are not fit can be either \textit{element} types or \textit{set} types. We start by looking at cases where the respondent is of a type \textit{element}.

Following the strategy we have defined, we are going to ask questions associated to sets that belong to $R$ until we get 1 as an answer. Lemma~\ref{lem:2} ensures that this interview would rule out all the \textit{element} candidates if it were asked to a \textit{null} candidate, and \textit{null} candidates answer 0 to all questions. Because of this, we can be sure that when we interview an \textit{element} candidate, they eventually answer 1. Otherwise, the interview would not rule them out if asked to a \textit{null} candidate.

As they must eventually answer 1, the \textit{null} candidates are ruled out. We will not rule out at least one \textit{element} candidate type (the respondent's type). As mentioned above, each question that is not \textit{private} rules out exactly one \textit{set} candidate type. Therefore, the interview  rules out $k'$ \textit{set} candidate types, where $k'\in \mathbb{N}$ and $k'\leq k$.

After the interview, the ratio of candidates with the private characteristic in the remaining population will be $(|S| - k') / ((|U| - u')*\Omega + |S| - k')$, where $u'\in \mathbb{N}$ y $u'<|U|$. We need to prove that $(|S| - k') / ((|U| - u')*\Omega + |S| - k')\in [a, b]$:

\begin{equation}
(|S| - k') / ((|U| - u')*\Omega + |S| - k') \geq (|S| - k') / (|U|*\Omega + |S|)
\end{equation}

\begin{equation}
(|S| - k') / (|U|*\Omega + |S|) \geq (|S| - k) / (|U|*\Omega + \Omega^2 + |S|) = a
\end{equation}

As $b=1$ and trivially the ratio of candidates with the private characteristic cannot be greater than 1, the value is in $[a, b]$.

The ratio of fit candidates in the remaining population is 0, because we have ruled out all the fit candidates and $0\leq x$. When the respondent is of type \textit{element}, the transformation is correct for positive instances.

Having seen this, it is easy to study the cases where the respondent is of a \textit{set} type. Candidates of one of these types answer 1 to the private question and only one other question. Therefore, they can answer 0 to either all the interview questions or to all but one.

If the respondent answers 1 to any question, then we are in a situation equivalent to the one where the respondent was of an \textit{element} type. The only difference is that it is now possible to rule out all \textit{element} candidates; instead of $u'<|U|$, $u'\leq |U|$. This change does not affect the previous argument.

If the respondent answers 0 to all questions, then we are in a situation equivalent to where the respondent was of the \textit{null} type. We rule out the same candidates and the ratios of fit candidates and candidates with the private characteristic are the same. As mentioned in the example seen in Section~\ref{sec:transformation example}, in this case the fitness prediction of the interview is incorrect, but the definition of \DCDP\ allows us a margin of error.

Thus, we have demonstrated that if the instance of \SCP\ is positive, then the transformed \DCDP\ instance is also be positive.

\subsubsection{Analysis for negative instances of the set cover decision problem}
\label{sec:negative instances}

In this case, we have to prove that if the result of an instance of \SCP\ is $\bot$, then the result of the instance of \DCDP\ arising from its transformation is also $\bot$.

First, we demonstrate that no correct interview without the private question can have more than $k$ questions. Trivially, all interviews of more than $k$ questions have to ask a $(k+1)$th question, where $k+1\leq |S|$. If we show that all interviews of $k+1$ questions infer more private information than allowed, then we will prove that all interviews of more than $k$ questions are incorrect.

For an interview to be correct, it has to be correct regardless of which candidate in the population is the respondent. Suppose the respondent is of type \textit{null}. As mentioned above, each question that is not \textit{private} rules out exactly one \textit{set} candidate type. In total, we rule out $k+1$ \textit{set} candidate types, no \textit{null} candidate types, and $u'\in \mathbb{N}$ \textit{element} candidate types, where $0\leq u' \leq |U|$.

Recall that the ratio of candidates with the private characteristic from the initial population is $|S| / (|U|*\Omega + \Omega^2 + |S|)$ and that $a=(|S| - k) / (|U|*\Omega + \Omega^2 + |S| - k)$ and $b=1$. After the interview, the ratio of candidates with the private characteristic from the remaining population will be $(|S| - k - 1) / ((|U| - u')*\Omega + \Omega^2 + |S| - k - 1)$. As $\Omega=2*|U|*|S|>0$, this ratio is less than $b$. To demonstrate that it cannot be in $[a, b]$, we need to demonstrate that $(|S| - k - 1) / ((|U| - u')*\Omega + \Omega^2 + |S| - k - 1)  <  a$. It is known that:
\begin{equation}
|U|, |S| \in \mathbb{N}^+;\ \ \ \Omega = 2*|U|*|S|;\ \ \ 0\leq u' \leq |U|;\ \ \ 0\leq k<|S|
\end{equation}
The following holds because $|U|$ and $|S|$ are positive:
\begin{equation}
|S| < 2*|U|^2*|S| + 2*|U|^2*|S|^2;
\end{equation}
We can add $2*|U|^2*|S|^2$ to both sides of the inequality and then perform several basic operations:
\begin{equation}
2*|U|^2*|S|^2 + |S| < 2*|U|^2*|S| + 4*|U|^2*|S|^2;
\end{equation}
\begin{equation}
(2*|U|^2*|S| + 1) * |S| < |U|*(2*|U|*|S|) + (2*|U|*|S|)^2;
\end{equation}
\begin{equation}
|S| < \frac{|U|*\Omega + \Omega^2} {|U|*\Omega + 1};
\end{equation}
If our inequality holds, the following must also hold:
\begin{equation}
|S| - k \leq |S| < \frac{|U|*\Omega + \Omega^2} {|U|*\Omega + 1} \leq \frac{|U|*\Omega + \Omega^2 + |S| - k} {|U|*\Omega + 1};
\end{equation}
\begin{equation}
|S| - k < \frac{|U|*\Omega + \Omega^2 + |S| - k} {|U|*\Omega + 1};
\end{equation}
\begin{equation}
\frac{|U|*\Omega + 1} {|U|*\Omega + \Omega^2 + |S| - k}  <  \frac{1} {|S| - k};
\end{equation}
Note that $|U|*\Omega + 1 = (|U|*\Omega + \Omega^2 + |S| - k) - (\Omega^2 + |S| - k - 1)$:
\begin{equation}
\frac{(|U|*\Omega + \Omega^2 + |S| - k) - (\Omega^2 + |S| - k - 1)} {|U|*\Omega + \Omega^2 + |S| - k}  <  \frac{1} {|S| - k};
\end{equation}
\begin{equation}
1 - \frac{\Omega^2 + |S| - k - 1} {|U|*\Omega + \Omega^2 + |S| - k}  <  \frac{1} {|S| - k};
\end{equation}
We can add $(|U| - u')*\Omega$ to the numerator of the fraction that is being subtracted, preserving the inequality. We will then rewrite the right side of the inequality and continue to perform basic operations:
\begin{equation}
1 - \frac{(|U| - u')*\Omega + \Omega^2 + |S| - k - 1} {|U|*\Omega + \Omega^2 + |S| - k}  < \frac{(|S| - k) - (|S| - k - 1)} {|S| - k};
\end{equation}
\begin{equation}
1 - \frac{(|U| - u')*\Omega + \Omega^2 + |S| - k - 1} {|U|*\Omega + \Omega^2 + |S| - k}  <  1 - \frac{|S| - k - 1} {|S| - k};
\end{equation}
\begin{equation}
\frac{|S| - k - 1} {|S| - k}  <  \frac{(|U| - u')*\Omega + \Omega^2 + |S| - k - 1} {|U|*\Omega + \Omega^2 + |S| - k};
\end{equation}
Finally, we can conclude that the following inequality holds:
\begin{equation}
\frac{|S| - k - 1}{(|U| - u')*\Omega + \Omega^2 + |S| - k - 1}  <  \frac{|S| - k} {|U|*\Omega + \Omega^2 + |S| - k}
\end{equation}
It is shown that the ratio of candidates with private characteristics in the remaining population is outside $[a, b]$.

Furthermore, Lemma~\ref{lem:1} assures us that, if there is a correct interview that includes question \textit{private}, then there must be a correct interview that is identical to the original except that it does not contain \textit{private}. With this in mind, the only possibility left to consider that the instance of \DCDP\ resulting from the transformation of a negative instance of \SCP\ is positive is the case in which the interview has up to $k$ questions with none being \textit{private}.

Because of Lemma~\ref{lem:1}, by ruling out this case we would also be ruling out all cases that include the private question. As previously mentioned, an interview is incorrect if it infers private information or does not determine with sufficient accuracy the fitness of the respondent. If it fails for one single candidate then the interview is incorrect, even if for the rest of the candidates it works correctly. We show that an interview with up to $k$ questions with none being \textit{private} can never be correct by focusing on the case in which the respondent is of the type \textit{null}.

Lemma~\ref{lem:2} assures us that this interview cannot rule out all the \textit{element} candidates, since the original instance of \SCP\ is negative. It rules out no \textit{null} candidates and $k'\in \mathbb{N}$ \textit{set} candidates, where $0\leq k'\leq k$, because each question associated to a set of the original instance (all questions except \textit{private}) rules out exactly one \textit{set} candidate.

The ratio of fit candidates in the initial population is $\Omega^2 / (|U|*\Omega + \Omega^2 + |S|)$. After the interview, the ratio will be $\Omega^2 / ((|U| - u')*\Omega + \Omega^2 + |S| - k')$, where $u'\in \mathbb{N}$ and $u'<|U|$.

\begin{equation}
\Omega^2 / ((|U| - u')*\Omega + \Omega^2 + |S| - k') \leq \Omega^2 / (\Omega + \Omega^2 + |S| - k') =
\end{equation}
\begin{equation}
\Omega^2 / (2*|U|*|S| + \Omega^2 + |S| - k')
\end{equation}
And, as $0\leq k'\leq k \leq |S|$ and $|U|>0$:
\begin{equation}
\Omega^2 / (2*|U|*|S| + \Omega^2 + |S| - k') \leq \Omega^2 / (2*|S| + \Omega^2 + |S| - |S|) \leq
\end{equation}
\begin{equation}
\Omega^2 / (\Omega^2 + 2*|S|) <  \Omega^2 / (\Omega^2 + |S|) = y
\end{equation}

Finally, since this ratio is greater than $x=0$ (since $\Omega>0$), it is in $[x, y]$, so the interview does not bound the population to a subset with a sufficiently low ($[0, x]$) or high ($[y, 1]$) ratio of fit candidates and is incorrect. We have demonstrated that, if the instance of \SCP\ is negative, then the transformed instance of \DCDP\ is also negative.

\subsection{Membership in NP}
\label{sec:appendix NP}

A problem is in NP if, for each problem instance, given a solution assumption (\textit{certificate}) the validity of the solution can be verified in polynomial time. The certificates of \DCDP\ are going to be question trees. The answers of the respondents are deterministic, i.e. they always answer the same way to the same question, so the possible answers to a question divide the remaining population (the candidate types who match the respondent) into pairwise disjoint sets. Each path of the tree represents a succession of questions (vertices) and answers (edges), and each node represents the subset of candidates of the initial population that would answer the questions in the path from the root to the node in the way indicated by their respective answers.

For the rest of the proof of the NP membership of \DCDP\ and \GDCOP, let $n\in \mathbb{N}$ be the number of candidate types in the initial population and $m\in \mathbb{N}$ the number of questions in total.

For each node in a tree that certifies a solution to this problem, the remaining populations of its child nodes will be pairwise disjoint sets; the sets of the candidate types that would answer in the manner indicated by the path of the child node. Trivially, the union of these sets equals the remaining population of the parent node. By inductive reasoning, we can deduce that the remaining populations of the leaf nodes of any correct certificate are pairwise disjoint sets whose intersection is the initial population. Therefore, there can be no more leaves than candidate types ($n$). Figure~\ref{fig:ejemplo entrevista interactiva}, used previously as an example of an interactive interview, illustrates this property.

Since our definition of the problem does not admit repeated questions in the same path and there are $m$ different questions, the maximum depth of the tree will be $m$. Moreover, since there are as many paths as leaves, the tree has at most $n*m$ nodes. The size of any problem instance must have $n*m$ as a lower bound because the fitness and quantity functions have to be defined for all candidate types in the initial population, and each candidate type requires $m$ space to be represented (candidate types are total functions $\mathit{Question}\rightarrow \mathit{Answer}$ and there are $m$ questions). Therefore, the number of nodes in a tree is linear with respect to the size of the instance it certifies.

Having seen this, let us define a polynomial verification algorithm. We assume that we are verifying a tree of appropriate dimensions ---if this is not the case, then we know that it cannot be a correct certificate. This algorithm consists in traversing the tree, storing the answers of the path that is being processed. When reaching each leaf node, we have to determine which candidate types could have responded in the way that has been stored. To do this, we run through the answers that would be given to up to $m$ questions by the $n$ candidate types for which the function $g$ is defined. We can check whether the remaining candidate types of a leaf satisfy the fitness and privacy requirements that have been specified in the instance of the problem in linear time with respect to the number of remaining candidate types, simply by applying the fitness and quantity functions.

Note that it is not necessary to check at every node in the tree whether the privacy requirement holds. If all child nodes of a node satisfy it, then the parent also satisfies the privacy requirement. We demonstrate this property with a proof by contradiction. Suppose that all children of a node satisfy the privacy requirement. If the parent node did not satisfy it, then the ratio of candidates answering in a certain way to a private question in the parent would be outside a certain interval (depending on $a$ and $b$). However, since the children satisfies the requirement, we know that the ratio of candidates answering the same private question in the same way at the child nodes must be within the interval. Since the ratio of the parent's candidates is a weighted average of those of the children, it is impossible for the parent's to be outside the interval. We have reached a contradiction, and we can state that the property we wanted to prove is true. Therefore, by applying inductive reasoning we can see that it is sufficient to just check the privacy requirement at the leaf nodes.

Each leaf can be processed in $n*m$ time, and consequentely the verification algorithm in $n*m + n^2*m$ time. Let $t\geq n*m$ be the input size. Then, the time complexity of the verification algorithm is $\Theta(t^2)$.

Therefore, it has been demonstrated that \DCDP\ and \GDCOP\ belong to NP.

\end{document}